\documentclass[useAMS,usenatbib]{mn2e}
\input{psfig}
\usepackage{rotating}
\usepackage{natbib}
\usepackage[dvips]{epsfig}
\usepackage{afterpage}

\title[OH masers in M82]{High velocity resolution observations of OH 
main line masers in the M82 starburst}
\author[M.K. Argo et al.]{M.K.~Argo$^{1}$, A.~Pedlar$^{2}$, 
R.J.~Beswick$^{2}$, T.W.B.~Muxlow$^{2}$, D.M.~Fenech$^{3}$\\
1. Curtin Institute of Radio Astronomy, Curtin University of Technology, 
Bentley, Perth, WA 6845, Australia\\
2. University of Manchester, Jodrell~Bank Observatory, Macclesfield, 
Cheshire SK11~9DL, UK\\
3. Department of Physics and Astronomy, University College London, 
Gower Street, London WC1E 6BT, UK\\}

\def\farcs    {\hbox{$.\!\!^{\prime\prime}$}}
\def\kms      {\ifmmode {\rm km\,s}^{-1} \else km\,s$^{-1}$\fi}

\def\mjybmch  {mJy\,beam$^{-1}$\,channel$^{-1}$}
\def\rasec    {\hbox{$.\!\!^{\rm s}$}}
\def\degr     {\hbox{$^\circ$}}

\begin{document}
\maketitle

\begin{abstract} 
{
Using the VLA, a series of high velocity resolution observations have 
been made of the M82 starburst at 1.6\,GHz.  These observations follow 
up on previous studies of the main line OH maser emission in the central 
kiloparsec of this starburst region, but with far greater velocity 
resolution, showing significant velocity structure in some of the 
maser spots for the first time.  A total of thirteen masers were 
detected, including all but one of the previously known sources.  
While some of these masers are still unresolved in velocity, these 
new results clearly show velocity structure in spectra from several of 
the maser regions.  Position-velocity plots show good agreement with 
the distribution of H{\sc i} including interesting velocity structure on 
the blue-ward feature in the west of the starburst which traces the 
velocity distribution seen in the ionised gas.
}
\end{abstract}



\section{Introduction}

Masers are common in star forming regions within the Milky Way and 
many nearby galaxies are now known to contain masers of one flavour or 
another, with powerful megamasers detected out to greater and greater 
redshifts (e.g. the recent discovery of a water maser at a redshift of 
2.639; \citealt{impellizzeri08}).
Maser sources require dense molecular gas and a nearby source of energy 
to create the required population inversion.  The presence of one or 
more type of maser in a given region can therefore be used to probe the 
physical conditions of the gas, making them useful tools for 
understanding complex starburst systems.  

Galactic OH masers located in star forming regions are typically very 
compact ($\sim$10$^{13}$\,cm), have very narrow line widths 
($\sim$1\,\kms) and flux ratios (1667/1665) less than one.  In contrast, 
megamaser emission in ULIRGs tends to be spatially unresolved, have line 
widths $>$100\,\kms, flux ratios $>$1 and total intensities typically 
a factor of 10$^6$ greater that that of Galactic sources 
(\citealt{lo05}).  In between these classes lie the so-called 
``kilomasers" with luminosities around 10$^3$ timer greater than typical 
Galactic sources.  These are found in nearby starbursts where the star 
formation rate is intermediate between that of the Milky Way and those 
of distant ULIRGs.

The nearby starburst M82 is the prototypical starburst and one of the 
most intensively studied.  It is known to contain large amounts of gas 
(e.g. H{\sc i}, \citealt{wills00}; CO, \citealt{shen95}), have strong 
outflows (e.g. winds traced by H$\alpha$, \citealt{shopbell98}), 
numerous supernova remnants and H{\sc ii} regions (e.g. 
\citealt{muxlow94,mcdonald02,fenech08}) and a history of strong star 
formation due to a past encounter with M81, although there is evidence 
that the rate of star formation has decreased significantly in the last 
5\,Myr (\citealt{beir08}).

Various types of maser have been detected in the central starburst of 
M82 over the last few decades, as would be expected for a galaxy with a 
significant molecular gas content and strong continuum emission, and 
these features can be used to investigate the molecular gas conditions 
and dynamics within the galaxy.
Main line OH masers were first detected in M82 by \cite{weliachew84} 
and, while they are significantly stronger than maser emission seen in 
the Milky Way, they are fainter than the megamasers seen at the other 
extreme in AGN.

Previous studies of the OH masers in M82 (\citealt{argo07}) have found 
10 individual masing regions, none of which are spatially resolved with 
the 1\farcs4 beam of the VLA.  These observations, however, were 
designed to probe the OH absorption in M82, rather than the masers, so 
the spectral resolution was comparatively poor.

This paper describes follow-up observations carried out with the VLA in 
2006 at higher spectral resolution in order to investigate the maser 
population in more detail.
Unless otherwise stated, all positions are given in J2000 
coordinates in the format aa\farcs{\rm aaa} bb\rasec{\rm bb} corresponding 
to 09$^{\rm h}$55$^{\rm m}$aa\rasec{\rm aaa} and +69\degr40'bb\farcs{\rm 
bb} respectively.  At the distance of M82 (3.6\,Mpc, 
\citealt{freedman94}), one arcsecond corresponds to a linear distance of 
17.5\,parsecs.


\section{Observations}

High velocity resolution observations were carried out using the VLA in 
A configuration during February and March 2006.  These observations were 
intended to to enable the detection of masers with narrow velocity 
widths which were not bright enough to be detectable in the much wider 
channels of the 2002 observations, and attempt to resolve velocity 
structure in those already known.

\begin{table}
\begin{center}
\begin{tabular}{ccc}
ObsID	& Date		& Time (LST)	\\
\hline
AA302a	& Feb 7/8	& 0200 - 1030	\\
AA302b	& Feb 16/17	& 0230 - 1030	\\
AA302c	& Feb 21	& 0300 - 0900	\\
AA302d	& Mar 19	& 0330 - 0930	\\
AA302e	& Mar 27/28	& 0430 - 1230	\\
\hline
\end{tabular}
\caption{\label{table_runs}VLA observations carried out in 2006 during 
programme AA302.  The structure of each observation was similar and 
included scans on flux and bandpass calibrators as well as scans on 
M82 and 0954+745 at two frequencies designed to cover the required 
bandwidth.}
\end{center}
\end{table}

The experiment was carried out in five separate runs, details of which 
are given in Table \ref{table_runs}.  The sources 1331+305, 0319+415 and 
0954+745 were used as flux, bandpass and phase calibrators respectively.  
The observations each consisted of scans on the flux and bandpass 
calibrators as well as scans on M82 and 0954+745 at two frequencies such 
that the total frequency coverage was sufficient to cover the entire 
velocity range expected across M82 at both 1665 and 1667\,MHz with some 
overlap between IFs at the band edges.  To achieve this, the correlator 
was used in 4-IF mode with a bandwidth of 1.5\,MHz and 128 channels per 
IF, resulting in a velocity resolution of 2\,\kms.  Velocities in this 
paper are quoted relative to 225\,\kms, the systemic radio LSR velocity 
of M82.

The data from each run were edited and calibrated separately before 
being combined prior to imaging.  For each run, the process was as 
described below.
Firstly, the channel zero file (created by averaging the inner 75 per 
cent of the band) was examined for each source, polarization, IF and 
frequency in order to detect major problems such as dead antennas and 
bad scans.  This pseudo-continuum dataset was then calibrated using 
standard methods for continuum calibration with the VLA.  Each of the 
four IFs were calibrated separately.  Together with the flag 
table, the final calibration tables from this step, one for each 
IF, were then copied over to the corresponding line datasets.  A 
bandpass calibration was then performed for each frequency using 
0319+415 before the data were split out separately for each IF.  The 
datasets for each of the epochs were then combined to create one dataset 
per IF from which image cubes were made.  Each of the four cubes were 
then inspected separately using various tasks within AIPS.

When the first run, AA302a, was initially examined in the AIPS task {\sc 
possm}, a large amount of interference was discovered.  The data were 
examined more closely in the visual task {\sc spflg} and strong 
interference was found to be present on all sources, varying both in 
time and with baseline.  It is not telescope specific, nor is it 
apparent on all baselines at a given time.  It is generally stronger at 
1665\,MHz than in the 1667\,MHz data and both stronger and more 
prevalent in Stokes RR than LL, but largely it does only affect a small 
subset of the channels.
A patient examination of each source in each frequency, stokes and IF 
was made in {\sc spflg} for each run and data were flagged on all 
sources, polarisations, IFs and times where the problem was seen to 
appear.  Several checks were made throughout this process to ensure that 
all interference was removed.

The resulting calibrated cubes, one for each of the four IFs, were 
imaged and then continuum subtracted using a set of line free channels 
at either end of the band before being examined for maser features.  The 
mean $3\sigma$ noise in each cube was 2.7\,\mjybmch\ (similar 
sensitivity to the low velocity resolution MERLIN data from 1995 but a 
factor of almost six worse than the 2002 VLA observations due to the 
much narrower channels).  By comparing measurements of continuum sources 
in M82 it was found that, to within the uncertainties, the positions and 
flux scales matched in all four final datasets.

\begin{table*}
\begin{center}
\begin{tabular}{|cc|cc|cc|c|c|}
\hline
J2000 ID	& R.A.		& Dec.		& S$_{1665}$	& S$_{1667}$	& Ratio		& Velocity	\\
		& (J2000)	& (J2000)	& (mJy)		& (mJy)		& (1667:1665)	& (\kms)	\\
\hline 
53.64+50.1 	& 53\rasec643	& 50\farcs11	& 11.4		& 15.0		& 1.32		& 115		\\  
53.11+47.9 	& 53.113	& 47.88		& 20.2		& $<$2.7	& $<$0.13	& 47		\\  
52.73+45.8 	& 52.729	& 45.77		& $16.9$	& $<$2.7	& $<$0.16	& 3		\\  
{\bf 52.42+49.2} & 52.416	& 49.18		& 11.1		& $<$2.7	& $<$0.24	& 62		\\ 
51.94+48.3	& 51.937	& 48.29		& 11.5		& $<$2.7	& $<$0.23	& $-$7		\\  
{\bf 51.55+48.5} & 51.543	& 48.47		& $<$2.7	& 7.42		& $>$2.75	& $-$27		\\ 
51.27+44.3 	& 51.276	& 44.29		& 13.2		& 32.4 		& 2.44		& $-$113	\\ 
50.95+45.4 	& 50.950	& 45.35		& 29.1		& 34.4		& 1.18		& $-$133	\\ 
50.54+45.5 	& 50.542	& 45.50		& 5.15		& $<$2.7	& $<$0.52	& $-$34		\\ 
50.38+44.3 	& 50.381	& 44.24		& 22.8		& 116		& 5.09		& $-$151	\\ 
{\bf 50.07+43.1} & 50.066	& 43.06		& $<$2.7	& 7.66		& $>$2.84	& $-$158	\\ 
49.71+44.2 	& 49.714	& 44.21		& $<$2.7	& 12.1		& $>$4.48	& $-$83		\\ 
48.43+41.9 	& 48.432	& 41.96		& 6.18		& 19.0		& 3.07		& $-$125	\\ 
\hline
\end{tabular} 
\caption{\label{vla06Table}
Definite maser detections identified from the VLA 2006 data.  Limits on 
the measured $peak$ flux densities are given as 3$\sigma$.  Velocities are 
quoted relative to the systemic velocity of 225\,\kms.  Where a line 
was present in both frequencies, the measured velocities were the same 
in both frequencies.  The maser IDs here are constructed from the J2000 
positions measured in this dataset (relative to 09$^{\rm h}$55$^{\rm m}$ 
+69\degr 40') and are used throughout this paper.  Maser IDs in bold are 
those which are new in these observations.  Uncertainties on fluxes are 
$\pm$2.7\,mJy, and $\pm$1.1\,\kms\ on velocities.}
\end{center}
\end{table*}

\begin{figure*}
\includegraphics[width=14cm,angle=270]{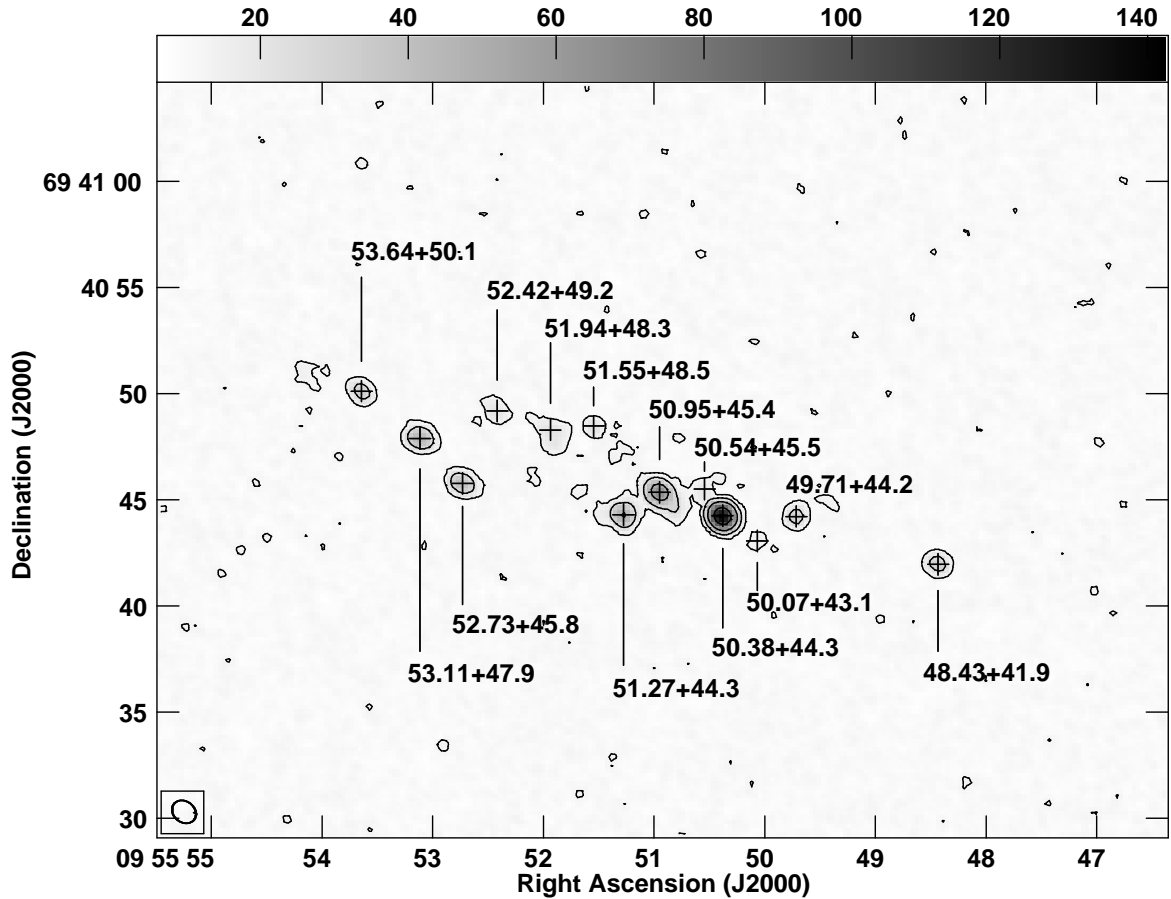}
\caption{\label{figmap}All masers detected in the VLA 2006 observations 
at $>$5$\sigma$.  The image is a combination of both frequencies showing 
masers at both 1665 and 1667\,MHz.  The contours are ($-$1,1,2,4,8,16,32) 
$\times$ 3\,mJy/beam and the labels correspond to the maser IDs given 
in Table \ref{vla06Table}.}
\end{figure*}

\section{Results and general comments}

Table \ref{vla06Table} lists all the maser features in this dataset 
which were present at 5$\sigma$ or more in more than one channel at 
either one or both frequencies, together with their 
positions, peak fluxes or limits in each line, flux ratio between the 
two lines and the velocity at the peak of the line profile.
Figure \ref{figmap} shows the maser locations across the central kpc of 
the M82 starburst.  This figure was constructed from the four final 
continuum-subtracted image cubes: each cube was collapsed in frequency, 
ignoring the five 
channels at either end of the band in each cube, and the resulting 
images were then summed to create one image.  The diagram shows 
all of the maser features above 5$\sigma$ in more than one consecutive 
channel, whether they are detected at 1665\,MHz, 1667\,MHz or both.

Despite the lower sensitivity per channel than in the VLA 2002 dataset, 
all but one of the previously known masers appear in this dataset.
Due to the narrower channels used in these observations, the masers 
generally appear much brighter here than in the 2002 dataset.  The 
exception to this is 50.02+45.8, a maser which has remained undetected 
since MERLIN observations in 1995 (see Section \ref{discussion}).

In general, fluxes measured in this dataset are more than a factor of 
two brighter than those measured in the much broader channels used in 
the 2002 observations.  The smallest difference is 50.95+45.4 where the 
difference is a factor of 1.9 at 1667\,MHz (2.78 at 1665\,MHz), and 
the largest is 53.11+47.9 which is brighter at 1665\,MHz by a factor of 
$>$10 in 2006.  This feature is narrow even at a velocity resolution of 
2\,\kms\ so could be brighter still.

Positions measured in the 2006 data are all within half a beam of 
those measured in 2002 at the same spatial resolution (1\farcs4).  The 
mean positional offset between sources in the two epochs is 0\farcs17 
with 51.94+48.3 having the largest offset (0\farcs36).

Spectra from each maser feature are shown in Figure \ref{1665spectra} 
(1665\,MHz) and Figure \ref{1667spectra} (1667\,MHz).  Most of the 
maser features have only one narrow peak, but a subset are broader 
with a shoulder on one side or the other.  The spectra are discussed 
further in Section \ref{discussion}.

\afterpage{\clearpage\begin{figure*}
\begin{center}
        \parbox{2.8in}{\psfig{figure=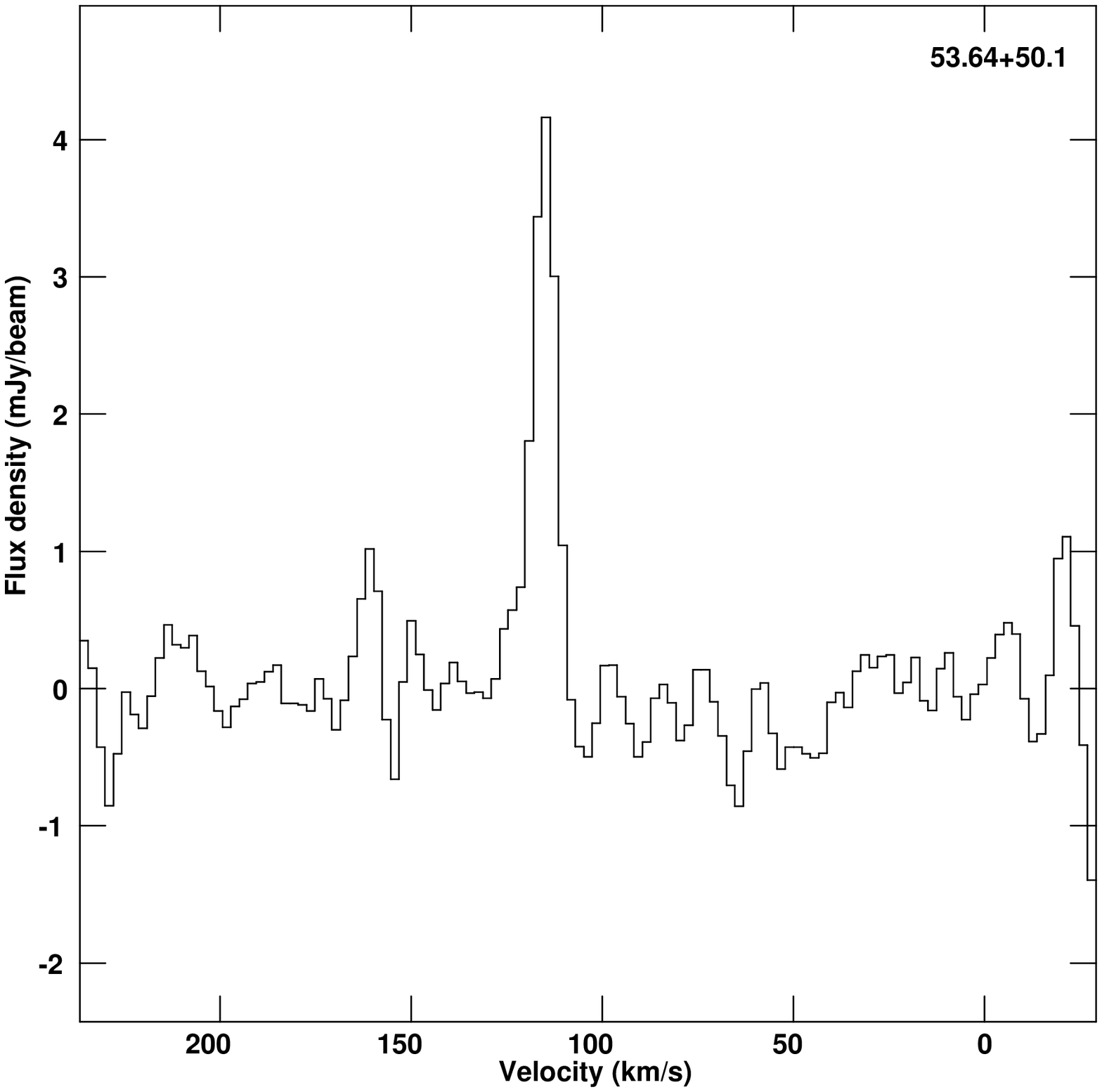, width=7cm}}
        \parbox{2.8in}{\psfig{figure=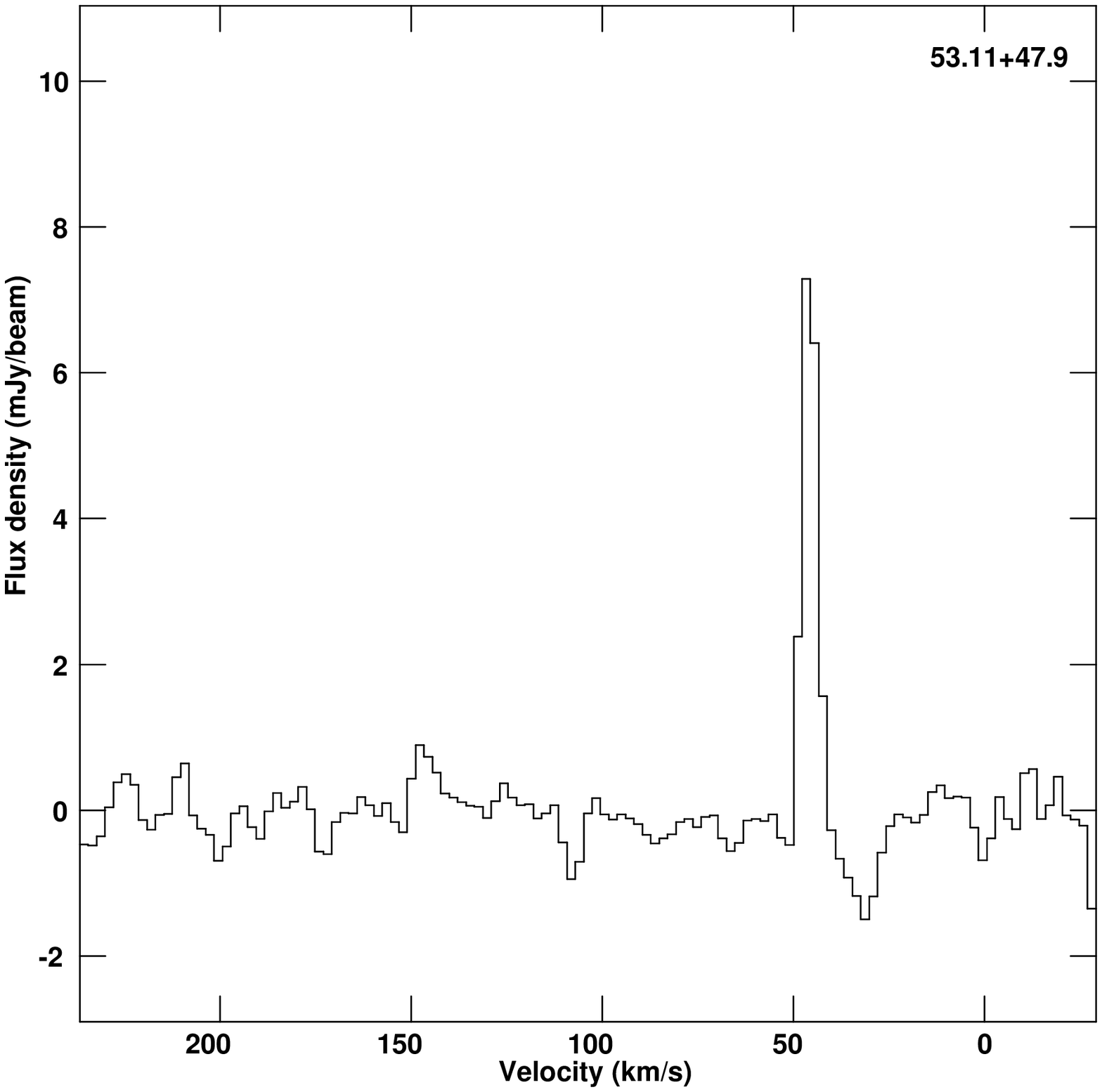, width=7cm}}
        \parbox{2.8in}{\psfig{figure=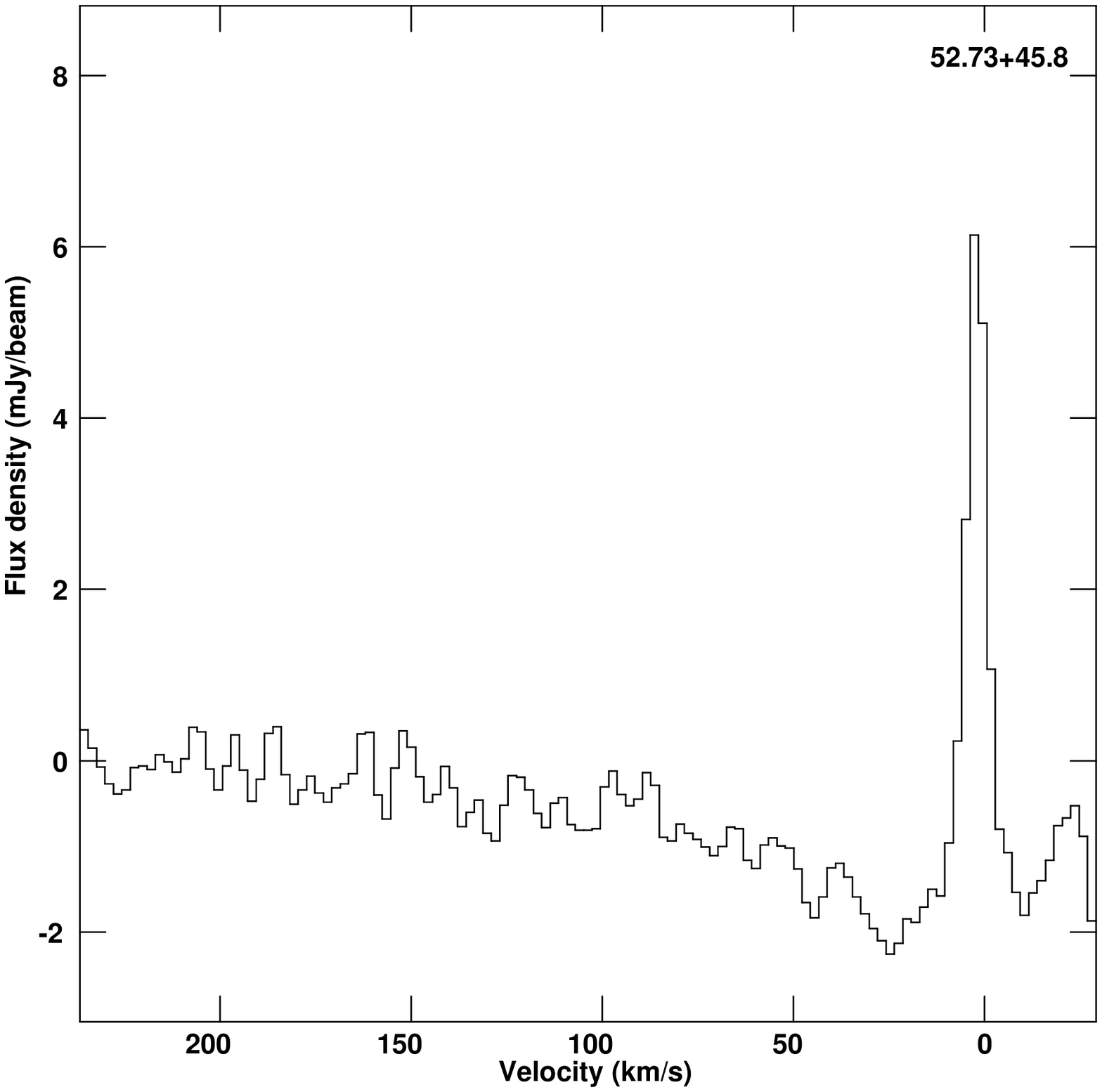, width=7cm}}
        \parbox{2.8in}{\psfig{figure=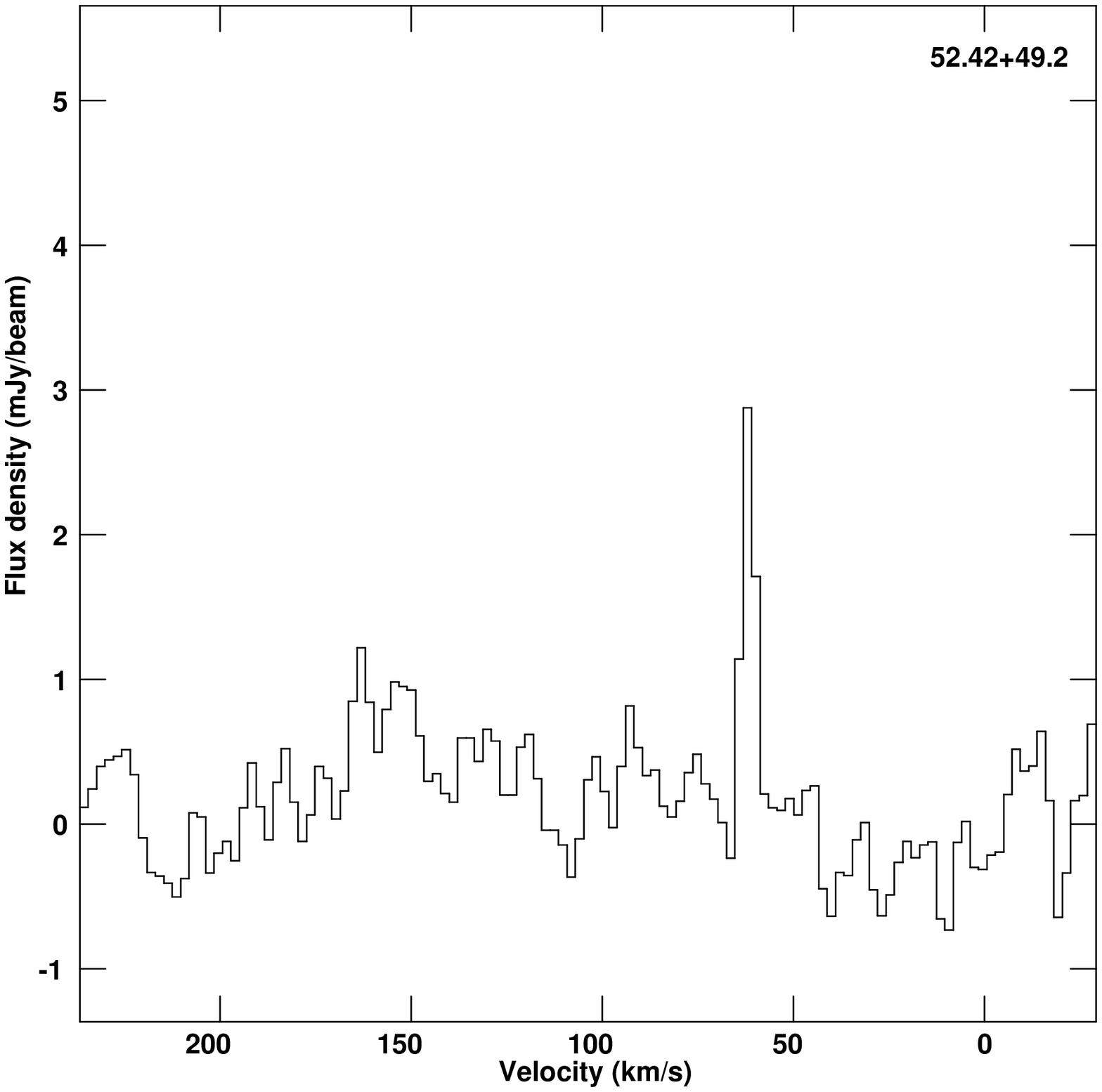, width=7cm}}
        \parbox{2.8in}{\psfig{figure=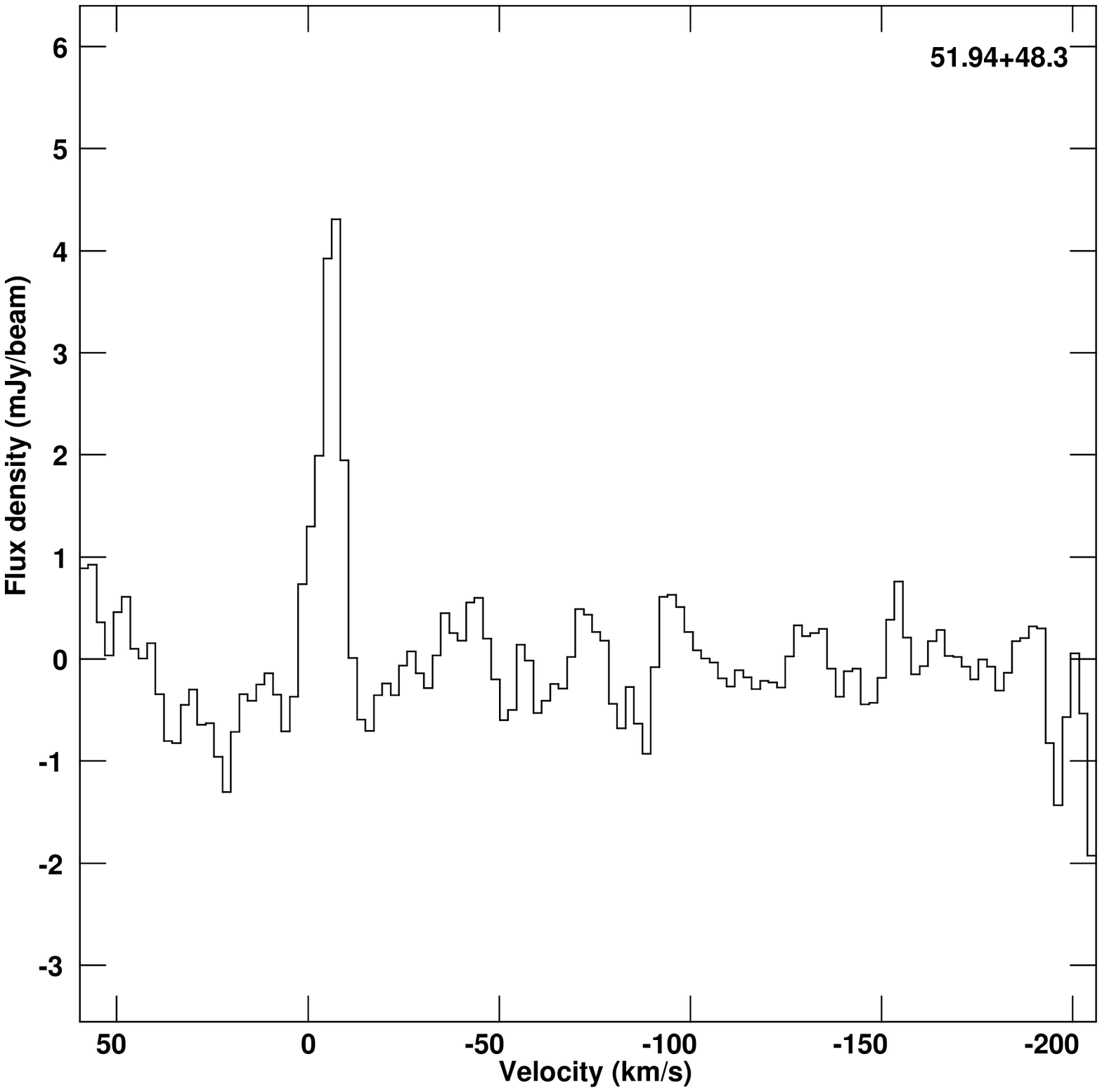, width=7cm}}
        \parbox{2.8in}{\psfig{figure=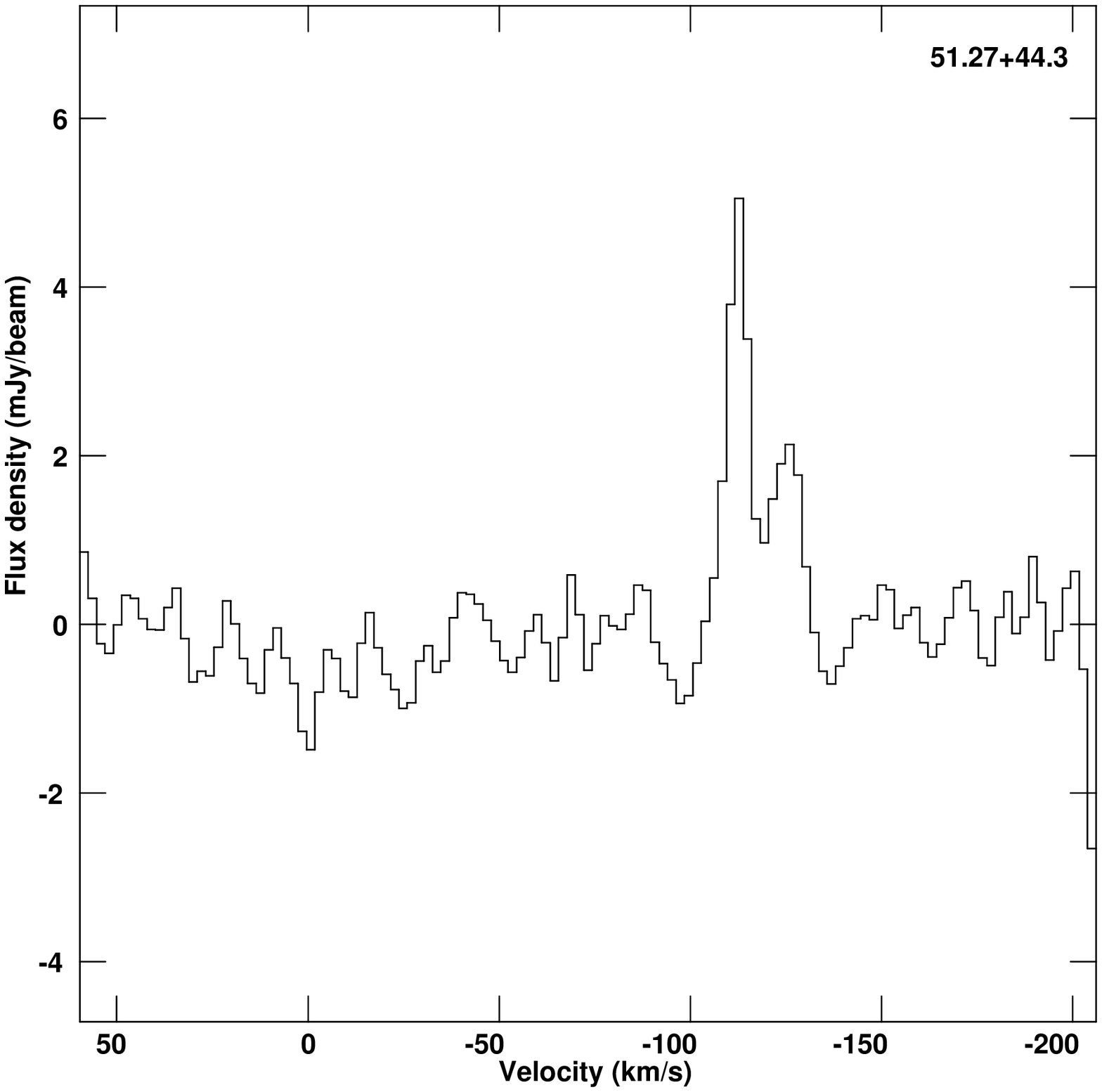, width=7cm}}
\end{center}
\caption{\label{1665spectra}Spectra of each maser listed in Table 
\ref{vla06Table} at 1665\,MHz, arranged in order of decreasing R.A.  
Each maser is seen in two or more adjacent channels with a flux greater 
than five times the rms noise level.  The data have been continuum 
subtracted and are smoothed with a Gaussian of width 2 channels.  The 
$x$-axis velocity scale is calculated relative to the rest frequency of 
the 1665\,MHz line.  Continued on next page.}
\end{figure*}
\begin{figure*}
\begin{center}
        \parbox{2.8in}{\psfig{figure=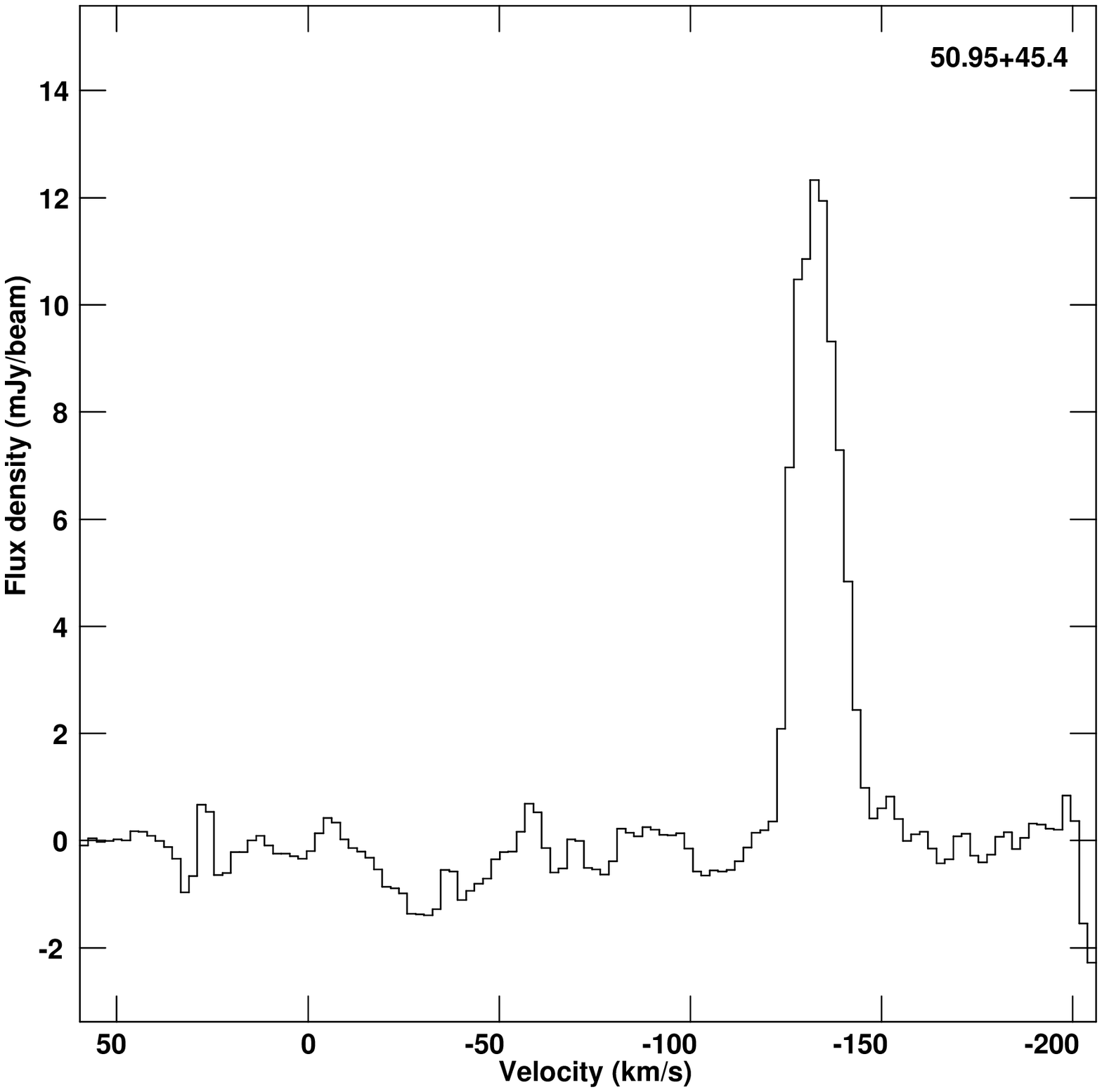, width=7cm}}
        \parbox{2.8in}{\psfig{figure=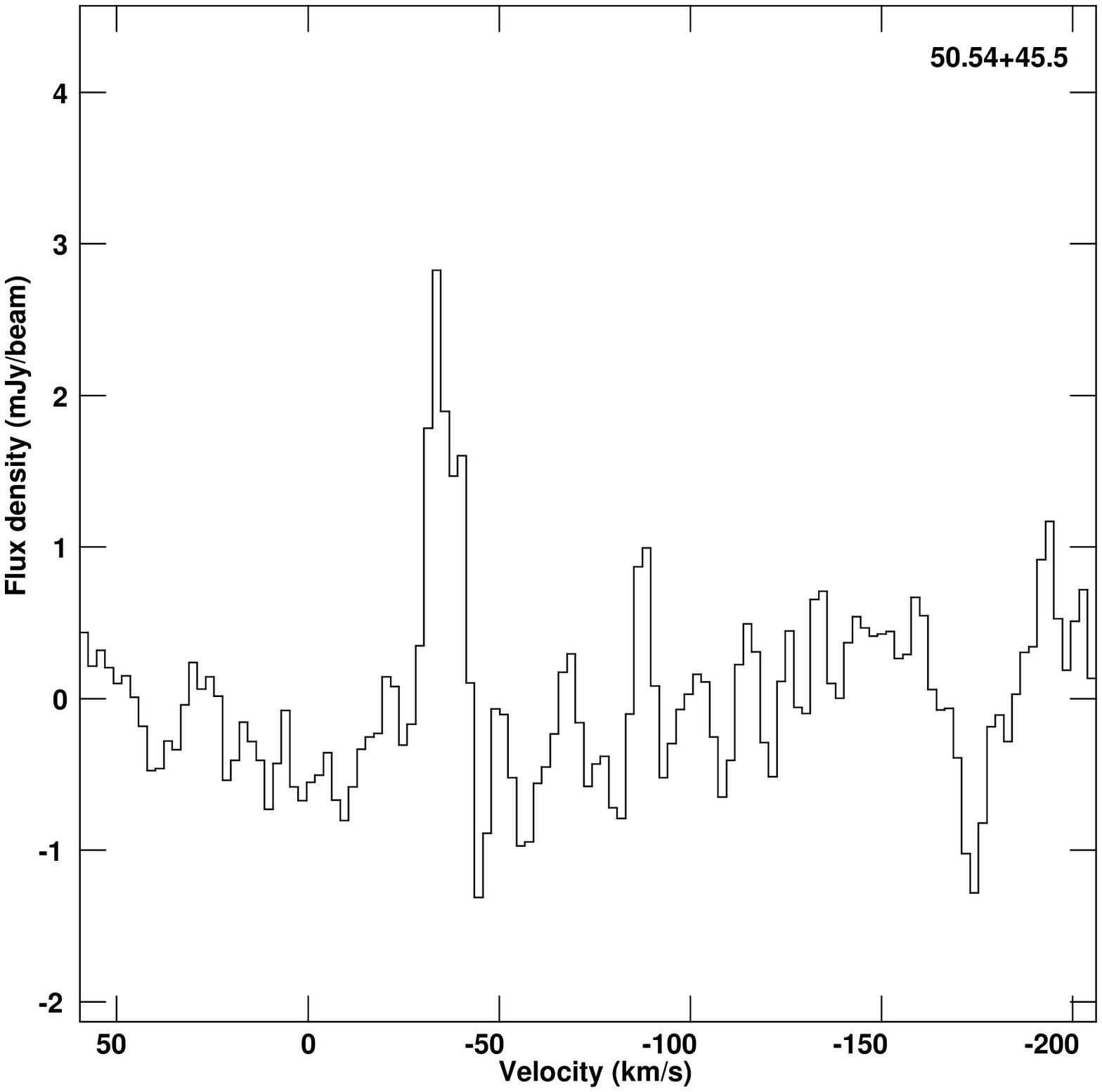, width=7cm}}
        \parbox{2.8in}{\psfig{figure=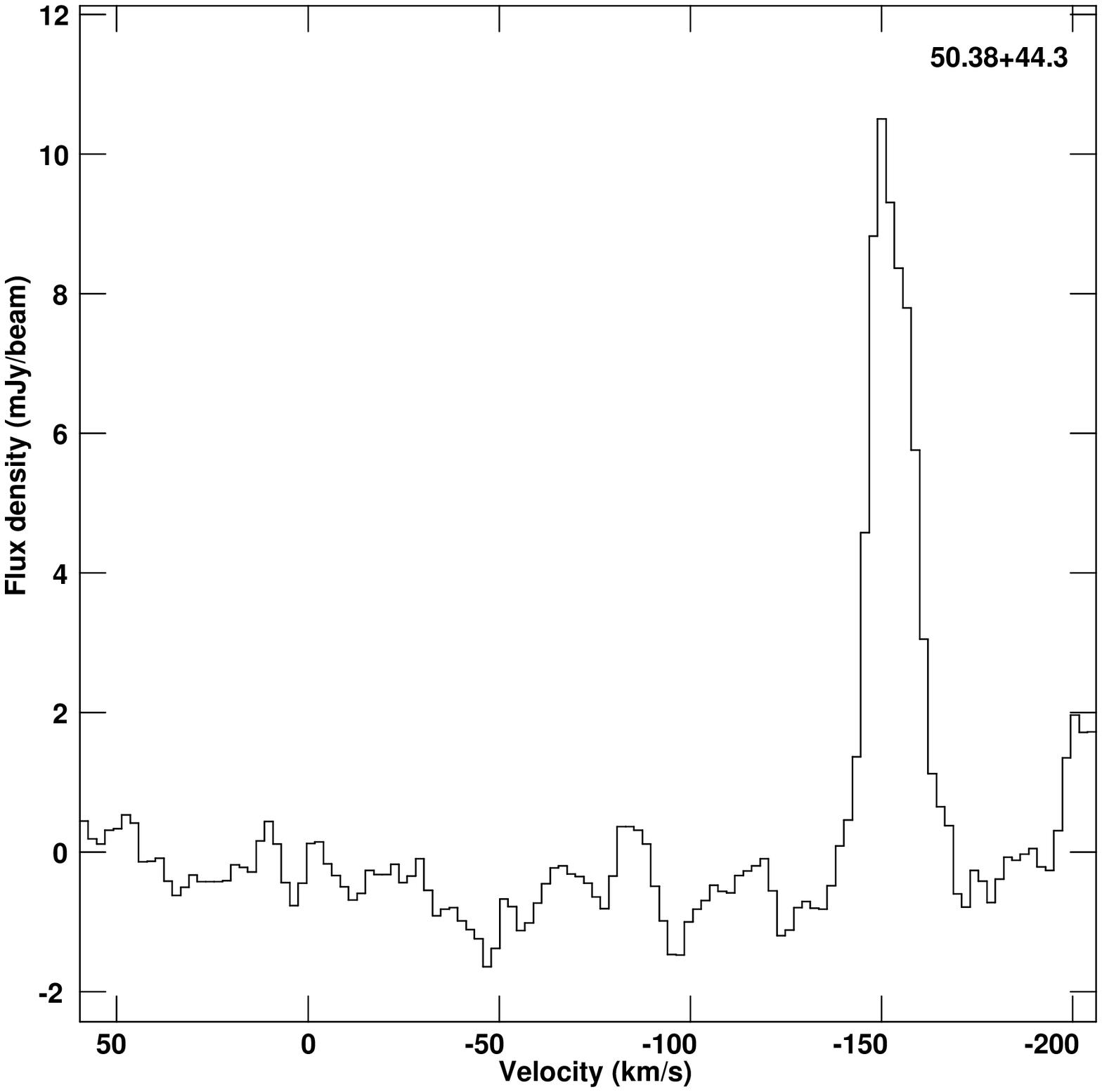, width=7cm}}
        \parbox{2.8in}{\psfig{figure=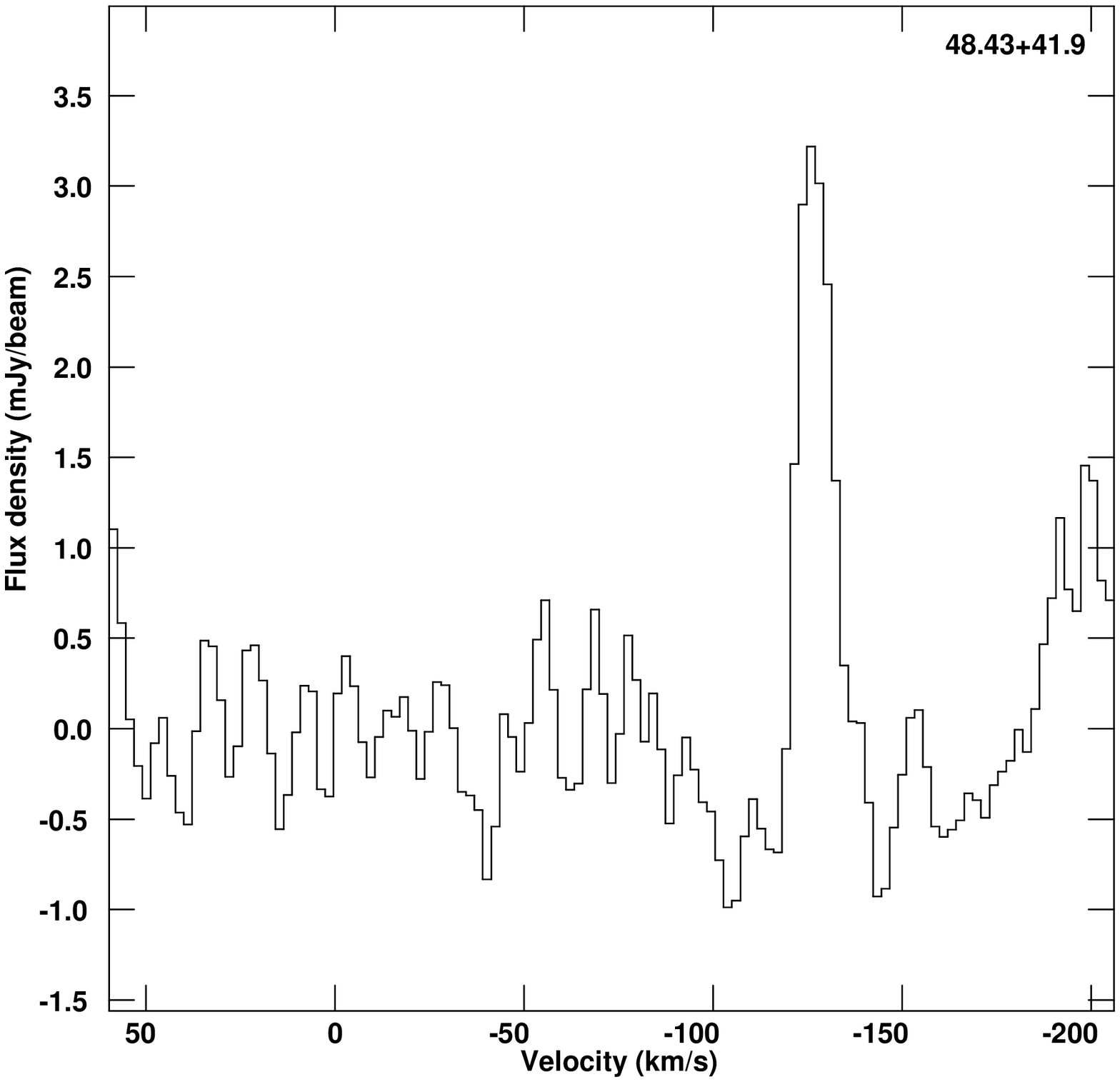, width=7cm}}\\
Figure \ref{1665spectra}: continued.
\end{center}
\end{figure*}
}
\afterpage{\clearpage\begin{figure*}
\begin{center}
        \parbox{2.8in}{\psfig{figure=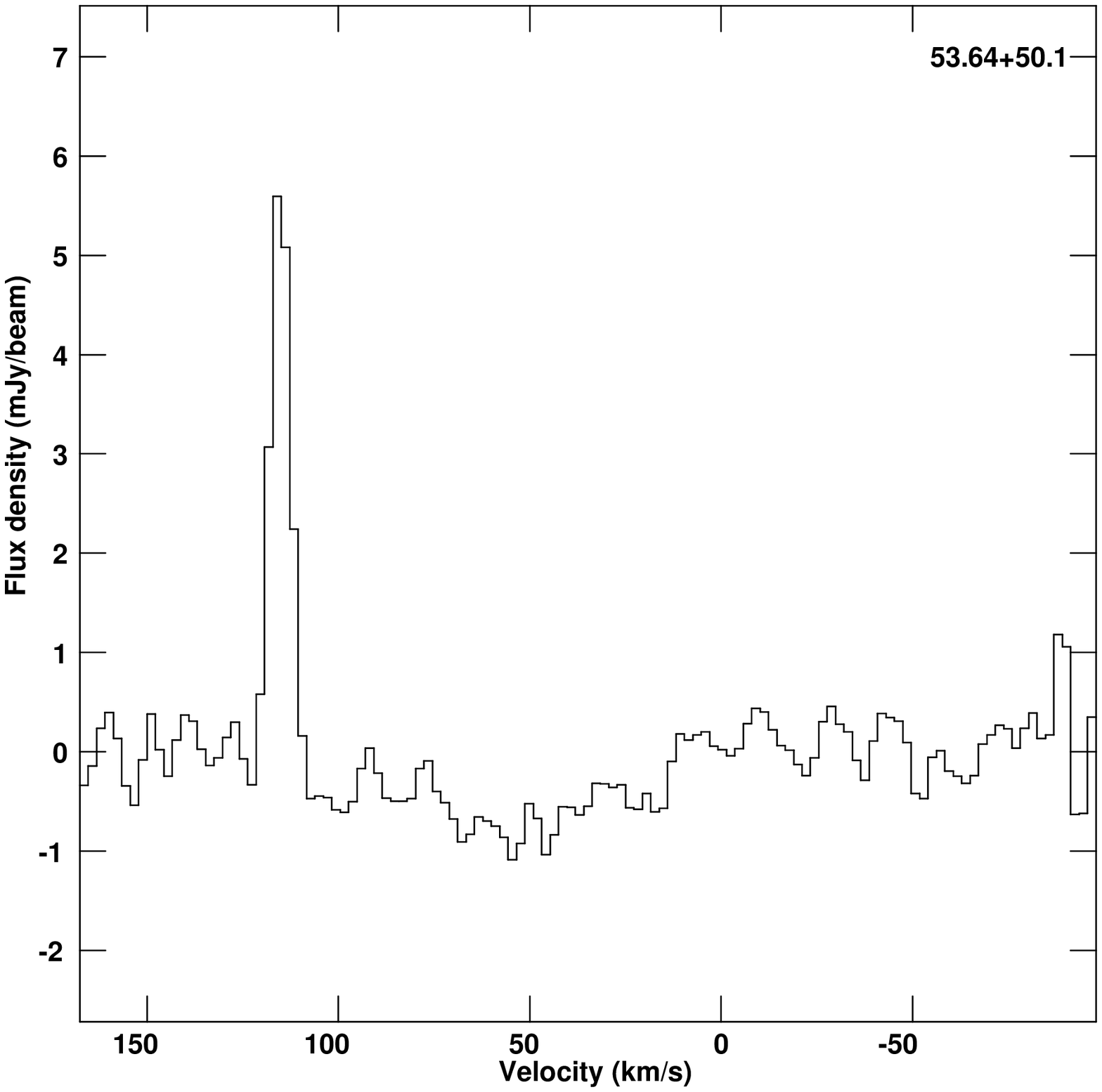, width=7cm}}
        \parbox{2.8in}{\psfig{figure=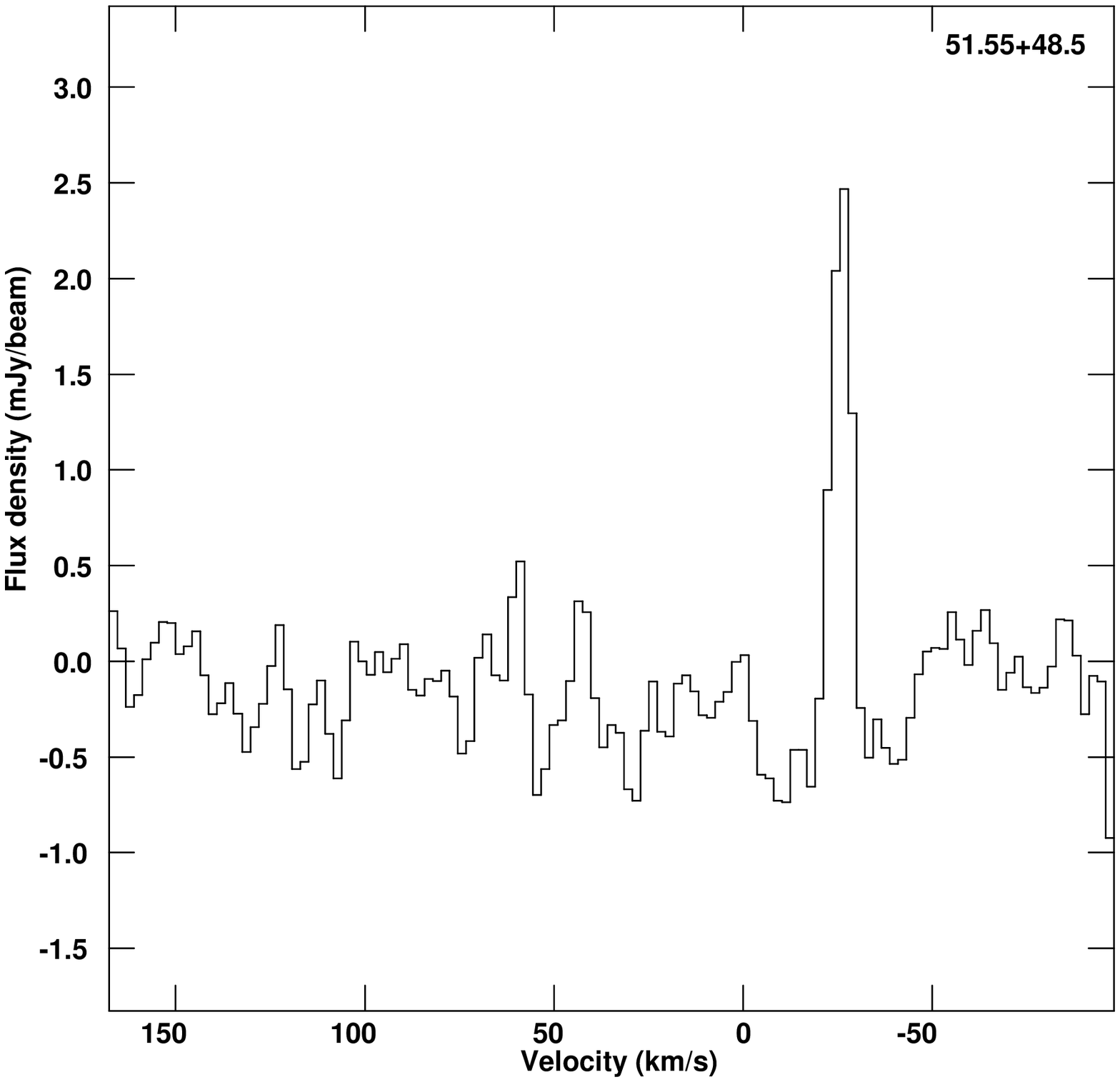, width=7cm}}
        \parbox{2.8in}{\psfig{figure=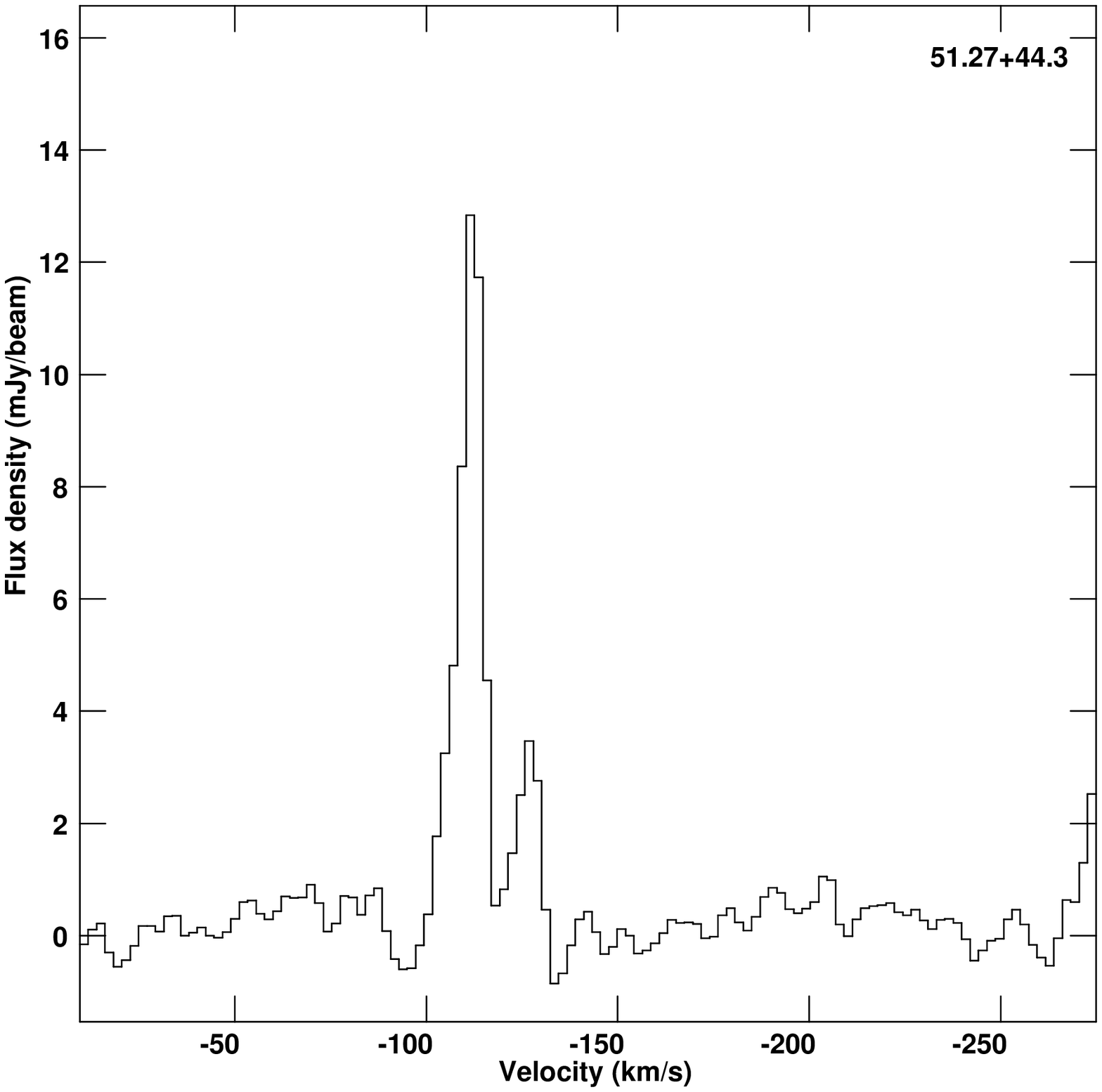, width=7cm}}
        \parbox{2.8in}{\psfig{figure=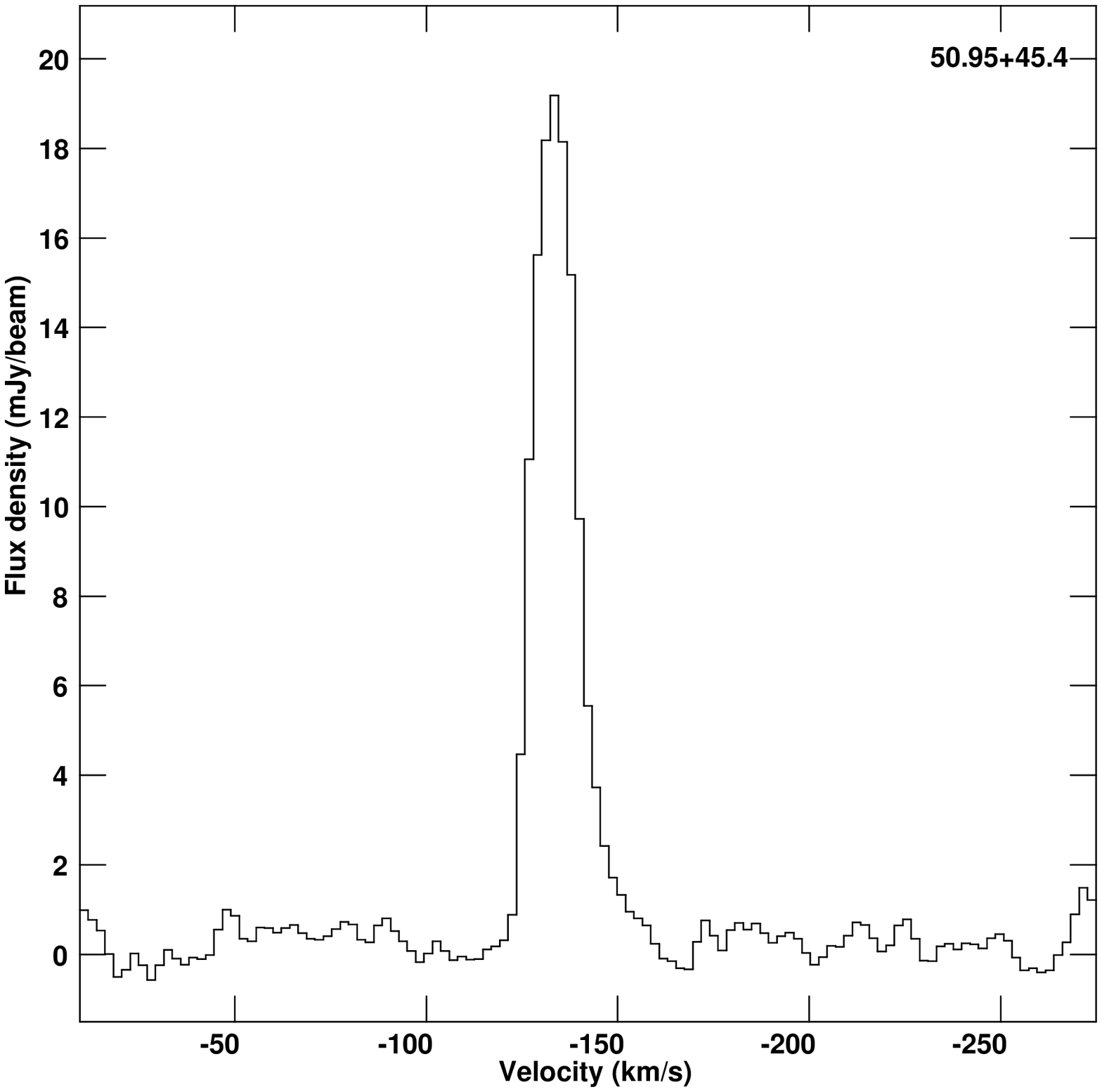, width=7cm}}
        \parbox{2.8in}{\psfig{figure=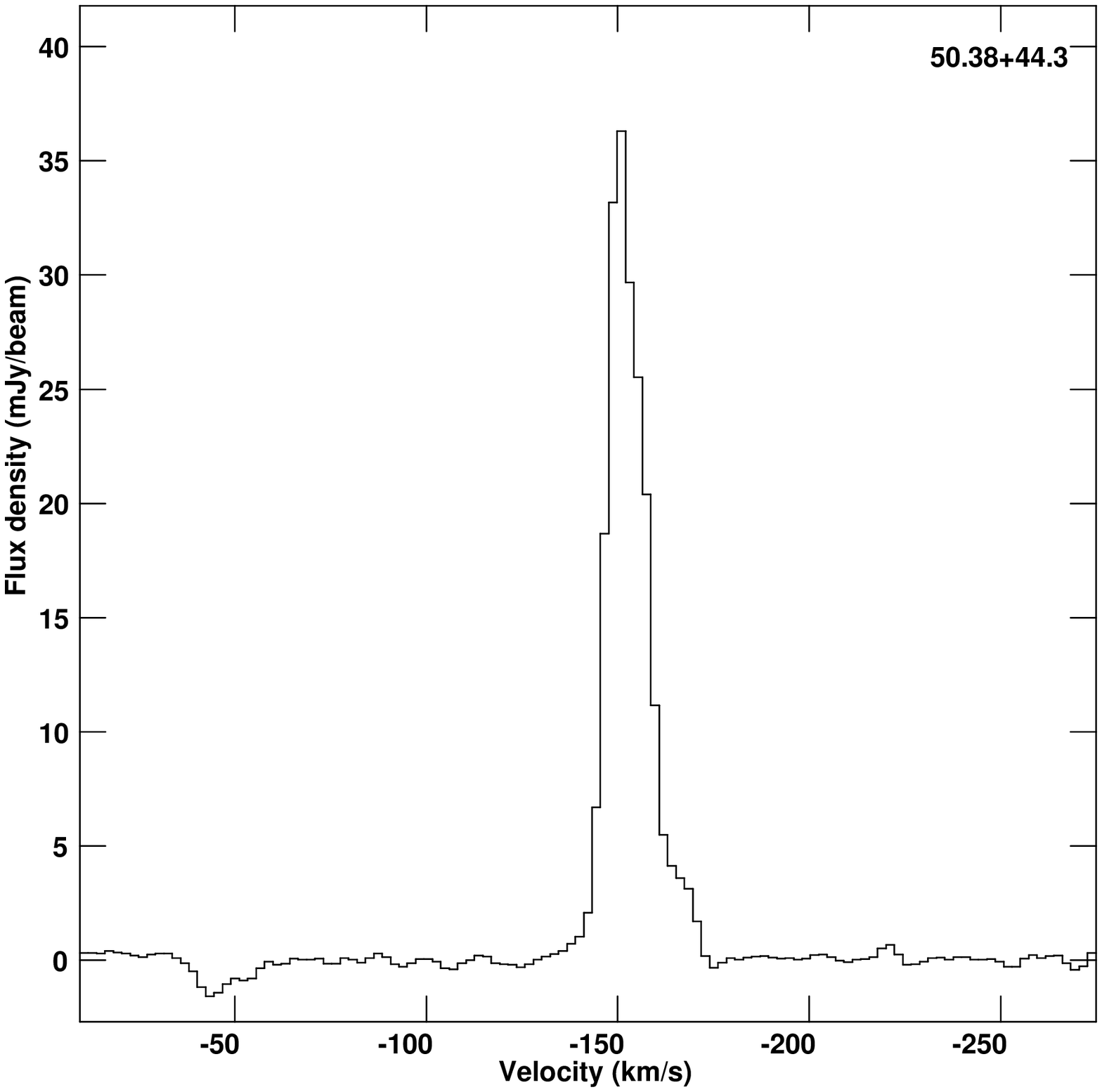, width=7cm}}
        \parbox{2.8in}{\psfig{figure=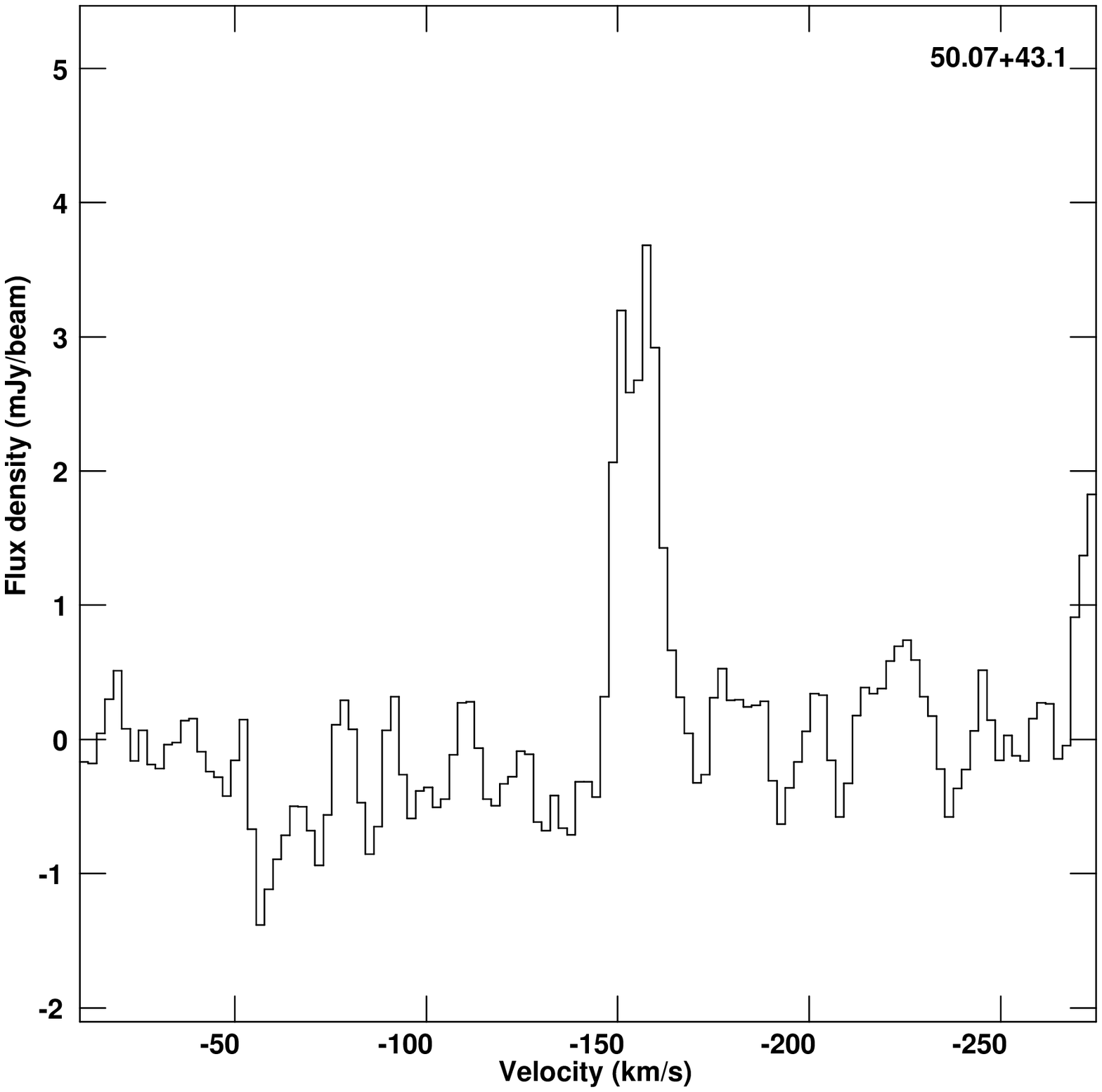, width=7cm}}
\end{center}
\caption{\label{1667spectra}Spectra of each maser listed in Table 
\ref{vla06Table} at 1667\,MHz, arranged in order of decreasing R.A.  
Each maser is seen in two or more adjacent channels with a flux greater 
than five times the rms noise level.  The data have been continuum 
subtracted and are smoothed with a Gaussian of width 2 channels.  The 
$x$-axis velocity scale is calculated relative to the rest frequency of 
the 1667\,MHz line.  Continued on next page.}
\end{figure*}
\begin{figure*}
\begin{center}
        \parbox{2.8in}{\psfig{figure=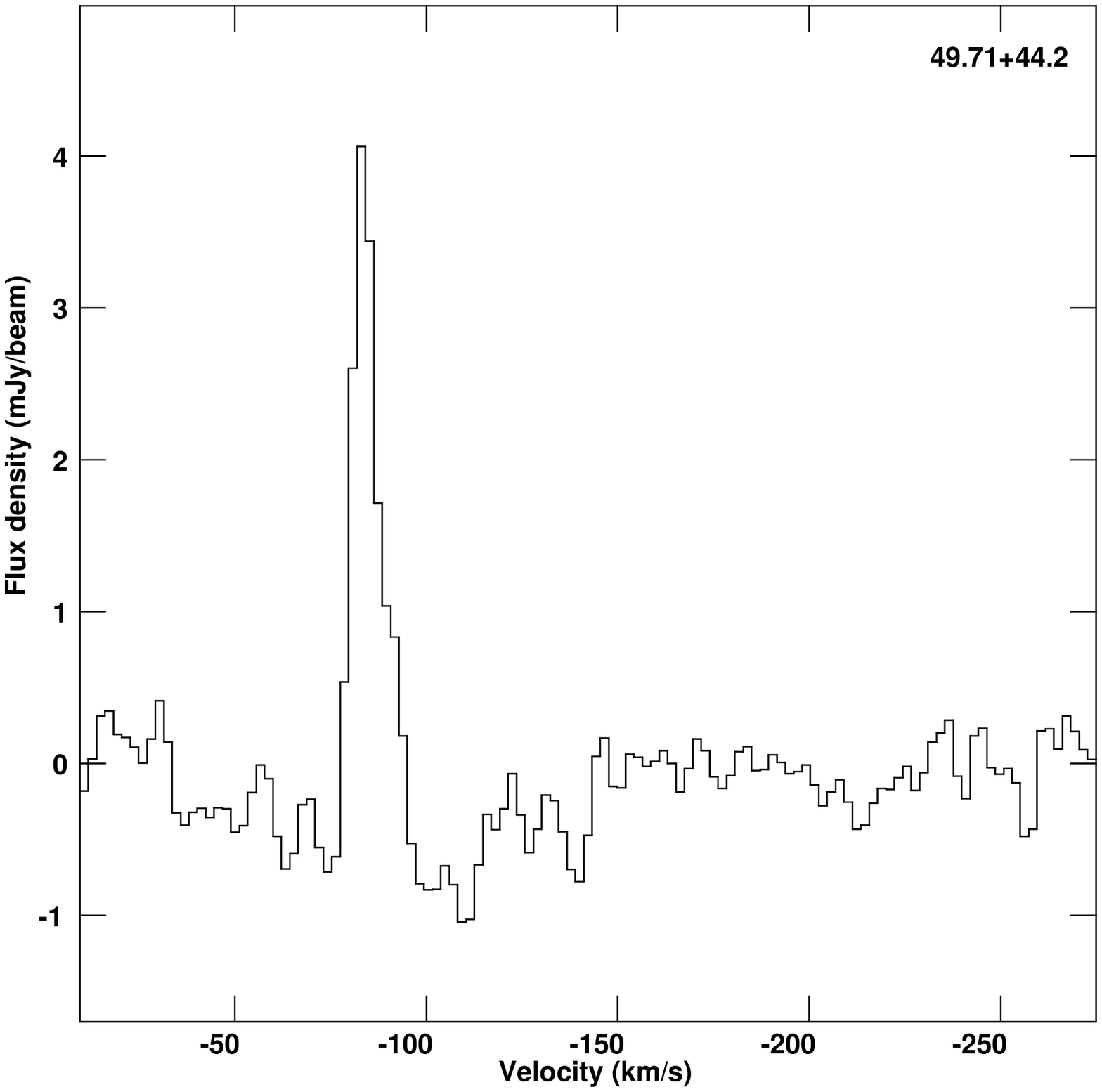, width=7cm}}
        \parbox{2.8in}{\psfig{figure=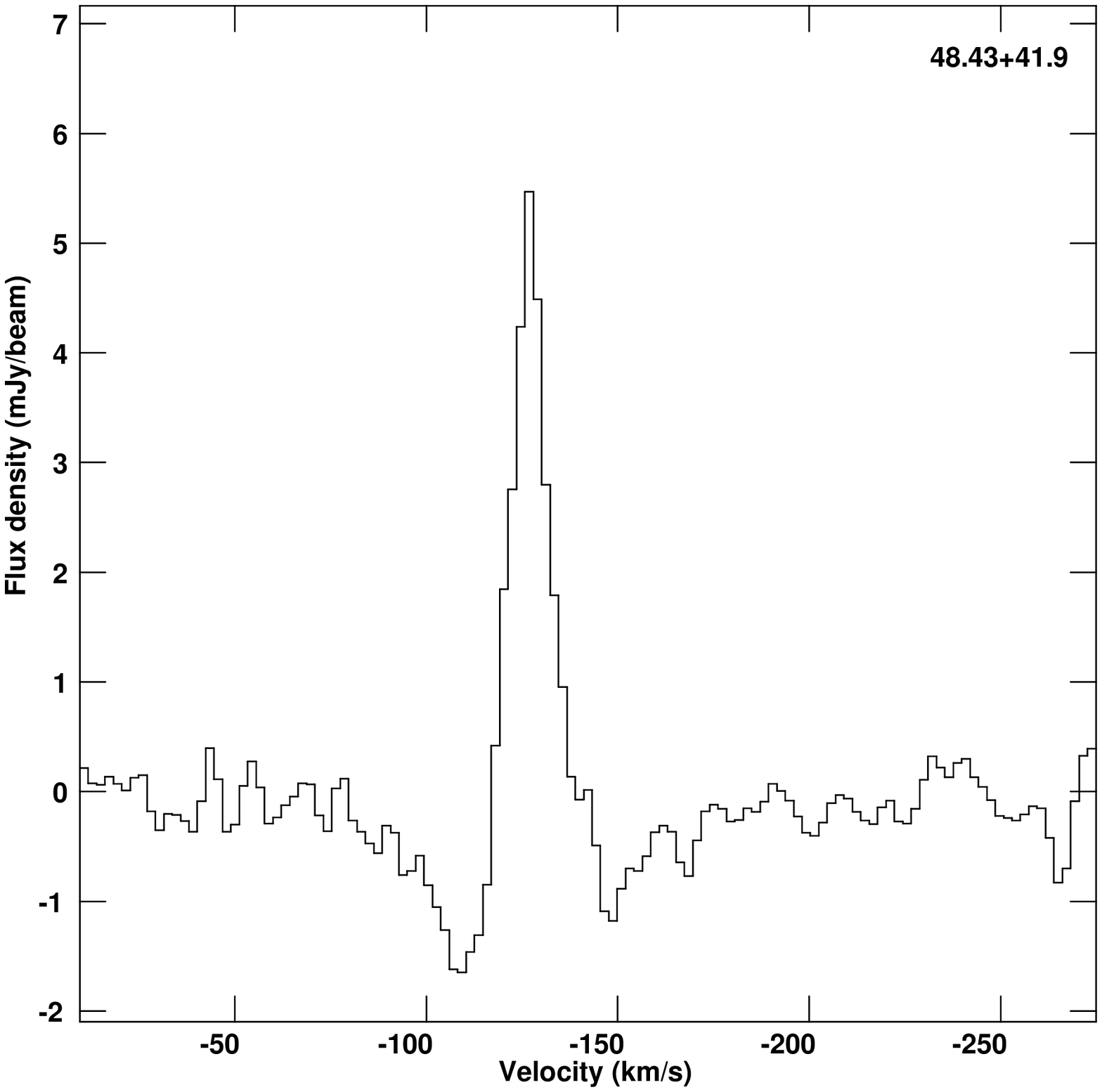, width=7cm}}\\
Figure \ref{1667spectra}: continued.
\end{center}
\end{figure*}
}

Although these observations were not ideal for detecting absorption, 
some is present in the data, mainly to the east of the dynamical centre.  
The deepest absorption features are coincident with those found in 
previous H{\sc i} studies at the eastern end of the central kpc.


\section{Notes on individual sources}

Brief notes are given here for each maser comparing with the results 
given in \cite{argo07}.  More detailed discussion of the properties of 
the population are given in the following section.
Note that the features are identified by their J2000 position in this 
dataset with labels corresponding to those in column one of Table 
\ref{vla06Table}.  Where a maser was detected in 2002, the velocities 
measured here are the same to within the uncertainties, with the 
exception of 51.94+48.3 (see below).  Unless otherwise noted, SNRs and 
H{\sc ii} regions are given by their B1950 positions as reported by 
\cite{mcdonald02}, H$_{2}$O masers are from \cite{baudry96} and OH 
satellite line features are from \cite{seaquist97}.

Note that all masers detected in the 2002 data are also detected here.  
The only previous maser which is undetected here is 50.02+45.8 which was 
only detected in the MERLIN 1995 observations.

(i) 53.64+50.1 is still detected in both lines but with a slightly 
larger line ratio than in 2002.  The measured velocity is the same 
(within the errors) for both lines and matches that measured in 2002 
(122$\pm$9\,\kms).  It lies within a beam of the H{\sc ii} region 
44.93+63.9 (B1950 position; \citealt{mcdonald02}).

(ii) 53.11+47.9 is again only present at 1665\,MHz where its velocity 
(46.7$\pm$1.1\,\kms) is consistent with that measured in 2002 
(56$\pm$9\,\kms).  It is apparently coincident with the supernova 
remnant 44.43+61.9.

(iii) 52.73+45.8 is also only present at 1665\,MHz, as in 2002, and 
the measured velocity is within the uncertainty of the previous result 
(8$\pm$9\,\kms).  This source is co-located with a supernova remnant 
(44.01+59.6) and a satellite line feature (F4; \citealt{seaquist97}) 
which is seen in absorption at 1720\,MHz and 
strongly in emission at 1612\,MHz with a velocity of 8$\pm$4\,\kms, 
consistent with the velocity measured here.  While listed as a supernova 
remnant by McDonald et al, the behaviour in the satellite lines led 
Seaquist et al to suggest the source as an AGN candidate, and EVN 
observations by \cite{wills99} show comparable structure and brightness 
to SgrA.
It is also within 1\farcs1 of an X-ray point source (CXOM82 
J095552.8+694047; \citealt{griffiths00}), but the main line ratio is 
atypical of megamaser sources associated with AGNs 
(\citealt{lonsdale02}).

(iv) 52.42+49.2 is a new maser in this dataset.  It is only present at 
1665\,MHz and is apparently coincident with a possible water maser 
detection (43.7+62.75) from \cite{baudry96}, listed as a possible 
feature only due to not being $>$5$\sigma$ in more than one channel.  
Interestingly, the authors note that this feature has an anomalous 
velocity compared to the molecular emission along the same line of 
sight.  The velocity measured for the OH feature 52.42+49.2 in the 2006 
data is $+$62.1\,\kms\ while the H$_2$0 feature of Baudry \& Brouillet 
is at a velocity of $-$64.6\,\kms, putting it well off the main 
distribution and away from any other features on the eastern side of the 
galaxy on the p-v diagram (see Section \ref{section_velocity}).

(v) 51.94+48.3 has the largest position offset compared to previous 
detections of all the masers discussed here, although the offset 
(0\farcs36) is smaller than the size of the beam (1\farcs4).  It was the 
weakest detection at 1665\,MHz in 2002 and not detected at all at 
1667\,MHz.  In 2006 it is also only detected at 1665\,MHz but with a 
slightly different velocity at it's peak (2002: 12$\pm$9\,\kms; 2006: 
$-$7.3$\pm$1.1\,\kms).  The detection in 2002 was only definite at 
1665\,MHz\footnote{Figure 4 of \protect\cite{argo07} shows the incorrect 
spectrum for this maser (labelled 51.87+48.3).} where it is superimposed 
on an absorption feature and hence was only just above the detection 
threshold.  The 2006 detection is much more significant.
It is along the same line of sight (within the size of the beam) as the 
continuum source 43.21+61.3 (B1950) from \cite{rodriguez04} detected at 
8.3 and 43\,GHz and classified as a possible H{\sc ii} region.

(vi) 51.55+48.5 is a new maser in this dataset.  It is only present at 
1667\,MHz where it is the weakest detection in this line and has no 
apparent associations with known continuum sources or other maser 
features.

(vii) 51.27+44.3 is detected in both lines as it was in 2002.  The 
measured position and velocity matches within the uncertainties (2002: 
$-$106$\pm$9\,\kms; 2006: $-$113$\pm$2\,\kms) but the line ratio has 
increased by a factor of two since 2002.  It is within a VLA beam of 
both a significant (S1) and a possible H$_{2}$O maser (42.5+59.25) from 
\cite{baudry96}, within an arcsecond of an H{\sc ii} region (42.48+58.4) 
and within a beam of two others.

(viii) 50.95+45.4 is once again also detected in both lines.  It is 
the only feature where the line ratio has decreased between the two 
epochs, but the velocities are the same within the uncertainties.  It is 
apparently coincident with both an H{\sc ii} region (42.21+59.2) and an 
H$_{2}$O maser (S2).

(ix) $50.54+45.5$ was brighter at 1665\,MHz in 2002 but was detected 
in both lines.  In the 2006 observations however, it is only detected 
(weakly) at 1665\,MHz.  It has no known associations with other masers 
or continuum features.

(x) $50.38+44.3$ is apparently coincident with a known H{\sc ii} region 
(41.61+57.9) and a water maser (S3).  It is again the brightest OH maser 
feature detected in M82 and continues to have the most extreme line ratio.

(xi) $50.07+43.1$ is a new feature in this dataset and is only detected 
at 1667\,MHz.  It is not coincident with any known continuum feature, or 
any other maser detection, and its velocity is consistent with it being 
located on the blue ward arc (see Section \ref{bluearc}).

(xii) $49.71+44.2$ is again only significantly detected at 1667\,MHz.  
This feature is within a beamwidth of two H{\sc ii} regions (40.95+58.8; 
40.96+57.9) and is also coincident with a possible H$_{2}$O feature 
(40.9+58.25).
At the same position there is a weak feature (3.7\,mJy) at 
1665\,MHz with a velocity ($-$79.8\,\kms) slightly offset from that 
measured at 1667\,MHz ($-$83.0\,\kms).  However, as it is $<$5$\sigma$, 
this 1665\,MHz feature is not included in further discussions.

(xiii) $48.43+1.9$ is again detected in both lines and is brighter at 
1667\,MHz with a velocity comparable to that measured in previous 
datasets.  It is within a beamwidth of an H{\sc ii} region (39.68+55.6) 
and an H$_{2}$O maser detection (S4).


\section{Discussion}\label{discussion}

Of the 13 features seen in these observations, ten have been 
identified in previous work (\citealt{argo07}).  Five masers are seen in 
both lines, a further five are seen only at 1665\,MHz and three are 
present only in the 1667\,MHz data.  The only maser reported previously 
which has not been detected again here is 50.02+45.8, a source which was 
only detected in low-velocity resolution MERLIN observations carried out 
in 1995 and, although it was detected at $>$3$\sigma$ in 1995, it was 
the weakest detection in that dataset and has not been detected 
subsequently.

This section discusses apparent associations, spatial structure, line 
ratios, spectra, velocity distribution and the blue ward arc in more 
detail.

\subsection{Possible associations}

Of the five masers visible only at 1665\,MHz, two are apparently 
coincident with supernova remnants (53.11+47.9; 52.73+45.8), one of 
which is also co-located with a satellite line feature (52.73+45.8), 
one lies within a beam of a possible H$_{2}$O feature (52.42+49.2), and 
two have no apparent associations (51.94+48.3; 50.54+45.5).  
Of the three sources seen only at 1667\,MHz, two are new detections 
(51.55+48.5; 50.07+43.1) neither of which appear associated with known 
continuum features or other maser detections.  The third of these, 
49.71+44.2, was previously noted to coincide with an H{\sc ii} region.  
The five features visible in both lines (53.64+50.1; 51.27+44.3; 
50.95+45.4; 50.38+44.3; 48.43+41.9) are all apparently coincident 
with known H{\sc ii} regions.

As has been noted previously, considering that the disk of M82 is 
viewed almost edge on these apparent associations are likely to be due 
to line of sight effects.  These observations lack the required spatial 
resolution to determine whether the sources are physically related.

\subsection{Spatial structure}
\newpage
\begin{figure*}
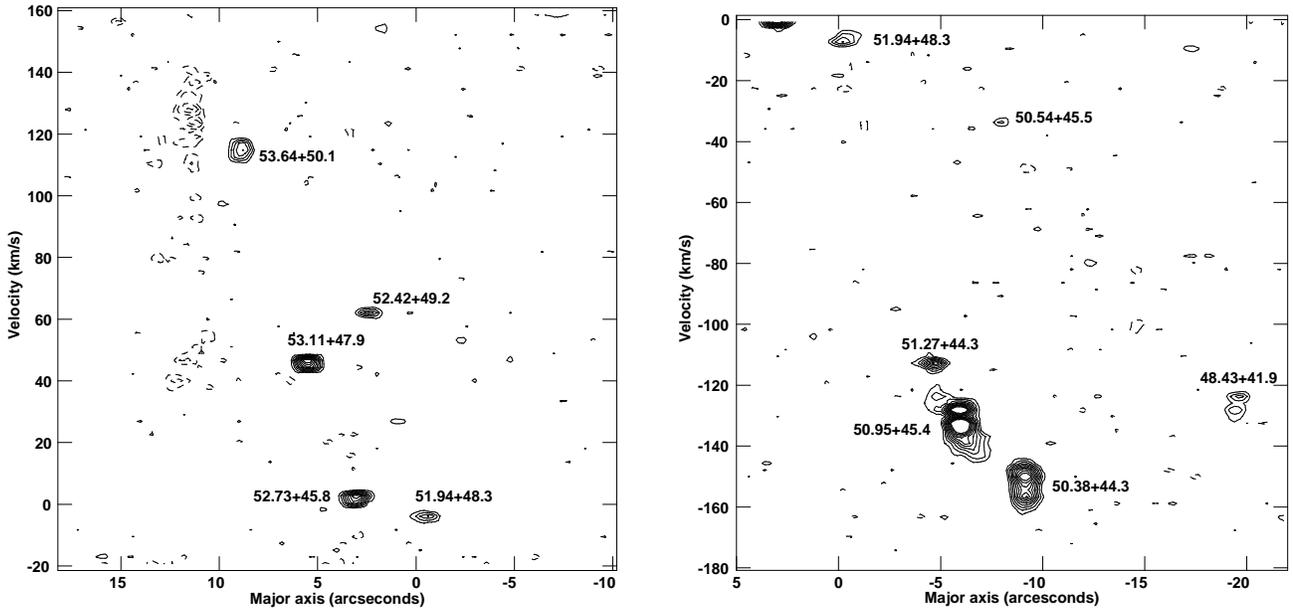

\begin{tabular}{cc}
\includegraphics[width=8.5cm]{FQ1_IF1_pv_kntr_nogrey2.ps}&
\includegraphics[width=8.5cm]{FQ1_IF2_pv_kntr_nogrey2.ps}\\
\end{tabular}
\caption{\label{pv1665}Position-velocity plots for each IF in the 
1665\,MHz dataset.  Features are labelled according to their positions 
as listed in Table \ref{vla06Table}.  Contours are ($-$10 to 10) 
$\times$ 1.6\,mJy/bm.  The velocity 0\,\kms\ represents the systemic 
velocity of M82 (225\,\kms) and the centre of M82 is taken to be 
09$^{\rm h}$55$^{\rm m}$52\rasec132 +69\degr40'46\farcs14 (J2000).}
\end{figure*}

\begin{figure*}
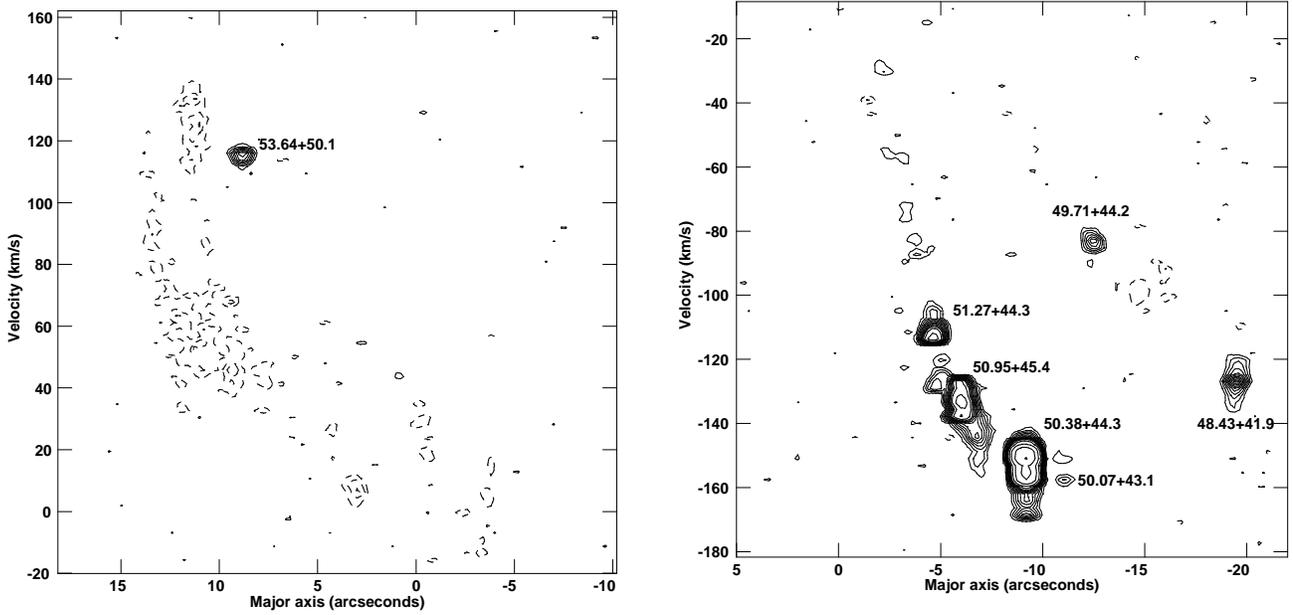

\begin{tabular}{cc}
\includegraphics[width=8.5cm]{FQ2_IF1_pv_kntr_nogrey2.ps}&
\includegraphics[width=8.5cm]{FQ2_IF2_pv_kntr_nogrey2.ps}\\
\end{tabular}
\caption{\label{pv1667}Position-velocity plots for each IF in the 
1667\,MHz dataset.  Features are labelled according to their positions 
as listed in Table \ref{vla06Table}.  Contours are ($-$10 to 10) 
$\times$ 1.6\,mJy/bm (left) and ($-$10 to 10, then 12, 16, 24, 40, 68) 
$\times$ 1.6\,mJy/bm (right).}
\end{figure*}

As with the 2002 VLA data, the resolution of these observations is 
insufficient to see spatial structure in the maser regions.  This is not 
surprising given that the physical size of typical Galactic masers 
placed at the distance of M82 would be much smaller than the linear 
resolution of the VLA at that distance (22\,pc at 1.6\,GHz).

The only maser which shows convincing physical extension is 50.95+45.4 
near the centre of the galaxy.  Previous VLA observations also showed a 
similar extension to the SW which was best fitted by three Gaussians. 
Higher resolution observations with MERLIN showed a weak second 
component along the same position angle (\citealt{argo07}).  These 
observations show an extension in the same direction (Fig. \ref{figmap}) 
which varies in frequency: the position of the peak moves from NE to SW 
as frequency increases.  
The same effect can be seen in both main lines and can also be seen 
clearly in the p-v diagram (see Section \ref{section_velocity}).
It is likely that the extension of 50.95+45.4 is due to several maser 
spots lying fairly close together but with different velocities, the 
resolution of the VLA at this frequency is not sufficient to distinguish 
them.

In the 1667\,MHz data, 51.27+44.3 appears to be slightly extended to the 
east.  Again, due to the small size of known Galactic maser regions, 
this is likely caused by the superposition in space of two unresolved 
maser spots, rather than any intrinsic structure.  This is also 
illustrated by features 50.54+45.5 and 50.38+44.3 which are located very 
close together on the sky, but are clearly separate features as their 
velocities are $-$34 and $-$151\,\kms\ respectively.

\subsection{Line ratios}

For an optically thin cloud in local thermal equilibrium, the expected 
ratio for emission from the 1665 and 1667\,MHz lines is 1.8.  Column 7 
of Table \ref{vla06Table} shows line ratios of each feature present in 
this dataset.  Where a feature is only detected in one line, the limit 
is calculated from the 3$\sigma$ noise in the other line.
Note that, compared to typical velocity resolutions used in observations 
of Galactic masers, these observations are still under-resolving the 
masers in M82, so it is possible that they could be brighter still.

In 2002, six features were detected in both lines allowing line 
ratios to be calculated.  One of these features, 50.54+45.5, had a line 
ratio of 0.69 in 2002 but is not detected above 5$\sigma$ at 1667\,MHz 
in 2006.  The other five masers in this group were all detected above 
5$\sigma$ in more than two channels in the 2006 observations in both 
lines.  Of these, three have line ratios which are consistent with those 
measured in 2002, given a typical 5 per cent flux uncertainty on each 
measurement.  However, 51.27+44.3 and 48.43+41.9 both have ratios 
significantly larger than those measured in the 2002 data.  In the case 
of 51.27+44.3 the ratio in 2006 is twice that measured in 2002.

As in 2002, 49.71+44.2 is again only detected at 1667\,MHz.  Two of the 
three new maser features (51.55+48.5 and 50.07+43.1) are also only 
present in the 1667\,MHz dataset and so have limits calculated from the 
3$\sigma$ noise measured in the 1665\,MHz line.
53.11+47.9, 52.73+45.8 and 51.94+48.3 are again only seen at 1665\,MHz 
in this dataset.  50.54+45.5 was previously detected at $>$5$\sigma$ in 
both lines in 2002 where it was stronger at 1665\,MHz, but in the 2006 
data it is only visible at 1665\,MHz.

With the exception of 50.95+45.4, for all masers which are detected in 
both lines the ratios have increased since the 2002 observations.  The 
greatest difference is for 51.27+44.3 which has doubled from a ratio of 
1.2 in 2002 to 2.4 in 2006.  The feature 50.38+44.3 continues to have 
the most extreme line ratio, 4.7 in 2002, 5.1 in 2006.
Interestingly, while it is a significant distance from the dynamical 
centre of M82 and apparently coincident with an H{\sc ii} region, 
50.38+44.3 has a high line ratio more typical of megamasers associated 
with AGN than Galactic masers in regions of star formation 
(\citealt{lonsdale02}).

One caveat with comparing measured line ratios between data sets is that 
whether or not the line is resolved affects the measured peak flux.  In 
some cases, the lines are clearly resolved in frequency, whereas with 
previous measurements they were unresolved.


\subsection{\label{section_velocity}Spectra and velocity structure }

Figures \ref{1665spectra} and \ref{1667spectra} show the spectra of each 
feature $>$5$\sigma$ in more than one consecutive channel at 1665 and 
1667\,MHz respectively.
The spectra are arranged in order of decreasing R.A. (left to right in 
Figure \ref{figmap}) and labelled 
according to their IDs in Table \ref{vla06Table}.
The p-v plots in Figures \ref{pv1665} and \ref{pv1667} show the 
distribution of masers in velocity along the major axis of the galaxy.  
Following the method of \cite{wills00}, before producing the p-v 
plots each cube was first rotated 17\degr\ anticlockwise so that the 
major axis of the galaxy was horizontal.  Then each cube was compressed 
to one plane along the declination axis, ignoring the five channels at 
either end of the band where the response was poor.

The velocity distribution of the masers in this dataset lies along the 
same distribution as that seen in the CO and H{\sc i} distributions in 
\cite{wills00}.  As seen in the 2002 observations however, several 
masers lie along an arc blue ward of the main distribution.  As with the 
H{\sc i} gas, the strongest features of this arc lie on the same 
axis as that of the ionised [Ne {\sc ii}] gas distribution of 
\cite{achtermann95}.  The cause 
of this feature has been suggested to be either an expanding superbubble 
or the $x_{2}$ orbits of an inner bar (\citealt{wills00}).

\begin{figure}
\includegraphics[width=8.5cm]{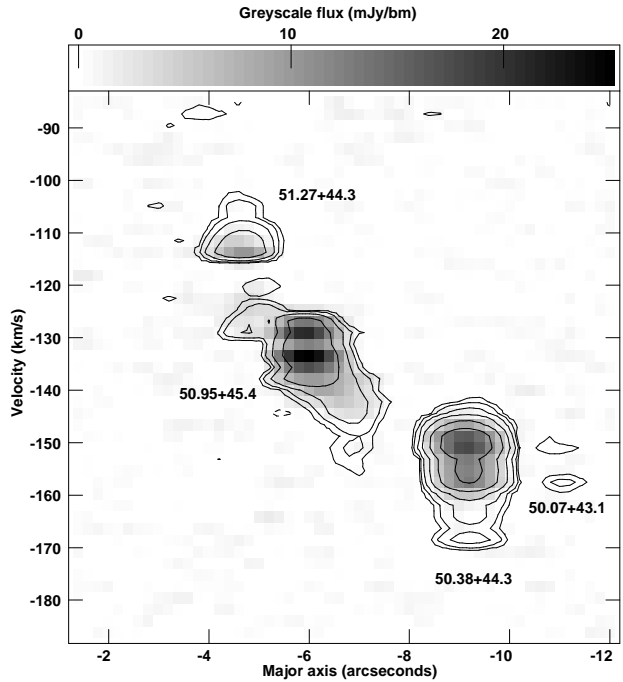}
\caption{\label{fig5097}The region around 50.95+45.4 in more detail.  
The 1665\,MHz data is shown as greyscale with a range from 0 to 25 
mJy/bm, while the 1667\,MHz data is plotted as contours of 
($-$1,1,2,4,8,16,32)$\times$1.9\,mJy/bm.  The emission at 1667\,MHz is 
noticeably broader in frequency than that at 1665\,MHz.}
\end{figure}

The spectra of the two brightest masers, 50.95+45.4 and 50.38.44.3, both 
show a blue wing at 1667\,MHz which is less obvious at 1665\,MHz.  It 
can be seen from the right-hand panels of Figures \ref{pv1665} and 
\ref{pv1667} that this region of the p-v diagram has some interesting 
structure.  Both 50.95+45.4 and 50.38+44.3 show velocity structure with 
significant extension in both lines, structure which is more extended at 
1667\,MHz than it is at 1665\,MHz.  This can be seen more clearly in 
Figure \ref{fig5097} where the two p-v plots are superimposed for this 
region.  The redward extension of 51.27+44.3 and the low-level bluest 
third peak of 50.38+44.3 are only visible at 1667\,MHz, while 50.95+45.4 
has broadly the same structure in both lines with extensions 
to both the east (red) and west (blue) and a double-peaked structure, 
with the extension again greatest at 1667\,MHz.

Another maser which shows velocity structure is 50.07+43.1 which appears 
to have two narrow peaks, but is only present at 1667\,MHz.  The two 
peaks here are more closely spaced in frequency than those of 51.27+44.3 
and have similar brightnesses, with the bluer peak marginally stronger.

In the p-v diagram, the feature 51.27+44.3 also clearly has two peaks at 
1667\,MHz with the bluer peak significantly stronger.  The p-v plots in 
Figures \ref{pv1665} and \ref{pv1667} show that there is also emission 
at the same position on the major axis but with an even bluer velocity, 
blending into the large feature at 50.95+45.4.  This illustrates the 
complexity of the ISM in M82.  The second, weaker peak in the spectrum 
of 51.27+44.3 in Figures \ref{1665spectra} and \ref{1667spectra} is 
actually this weak feature close to 50.95+45.4.  Given the linear 
resolution of the VLA, it is likely that this feature is a different 
maser spot, physically unconnected but along the same line of sight as 
51.27+44.3, with a velocity similar to that of 50.95+45.4.

The structure of this group of masers in the p-v diagram suggests 
several closely spaced maser regions sited along the blue ward arc, and 
it is likely that this region contains many clouds at different 
velocities spread out over a significant volume.  This region is 
discussed further in Section \ref{bluearc}.

The only source located in the main velocity distribution to show 
significant velocity structure is 48.43+41.9 which is weak but has a 
double-peaked profile in the p-v plot at 1665\,MHz, but a stronger 
single peak with pronounced wings at 1667\,MHz.


\section{Comparison with other wavelengths}\label{section_surveys}

Due to its proximity, M82 has been well studied across the 
electromagnetic spectrum.  It is interesting to compare the emission 
seen here in OH with features seen at other wavelengths in order to gain 
a greater understanding of the conditions and dynamics within the 
central kiloparsec of the galaxy.

\subsection{Stars and dust}

\begin{figure*}
\includegraphics[width=14cm,angle=0]{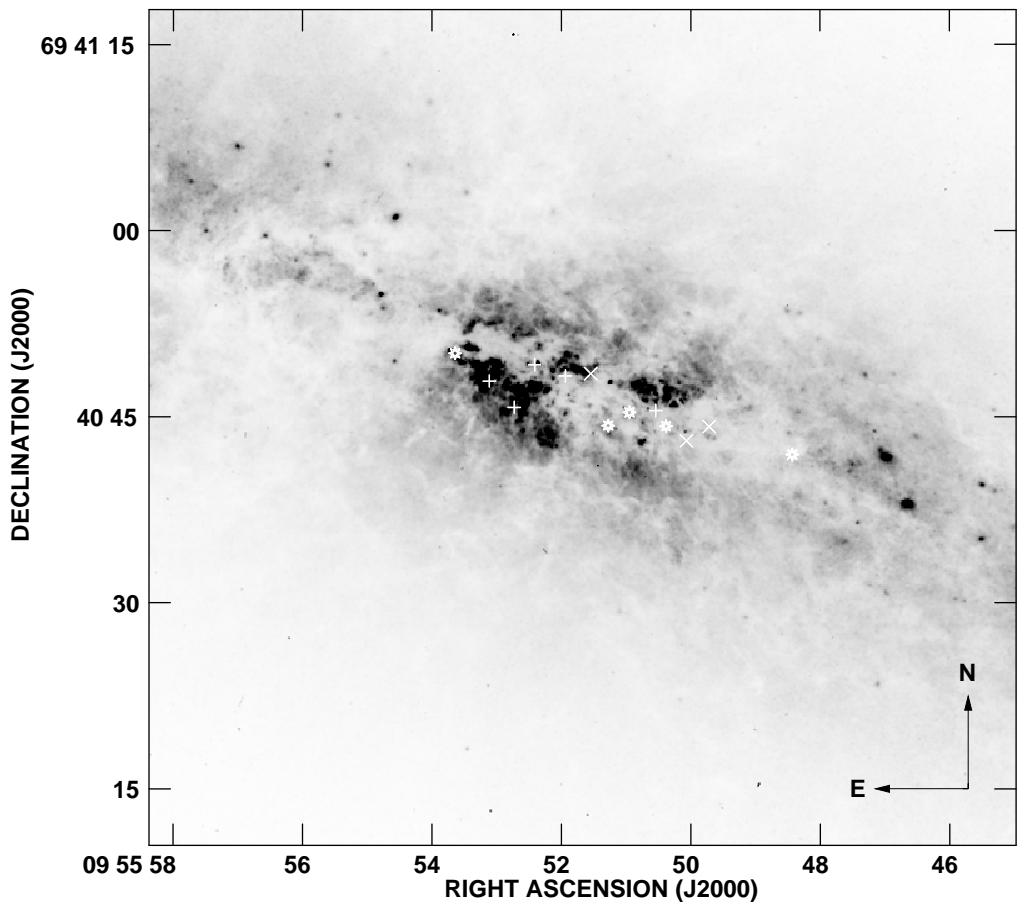}
\caption{\label{fighst}The OH masers in M82 plotted over the optical 
emission from the central starburst.  Grey scale: H$\alpha$ HST ACS image 
(program ID 9788, PI Ho).  The positions of the masers are also plotted.  
Those visible at just 1665\,MHz are shown as plusses, those detected 
only at 1667\,MHz as crosses, and those present in both lines as stars.}
\end{figure*}

Figure \ref{fighst} shows the positions of the masers plotted on an 
archival image from the ACS instrument on the $HST$ using a 680-980\,nm 
filter.  While it should be noted that there are a lack of directly 
comparable features between the optical and radio fields, the astrometry 
of the HST image was tweaked using positions of objects in the field 
from the 2MASS and SDSS catalogues, with a resulting estimated rms 
positional uncertainty of 0\farcs6 (the symbols are 1 arcsecond in 
diameter).  Different symbols indicate whether the 
maser is detected at just 1665\,MHz, just 1667\,MHz, or both.  As this 
figure shows, and as might be expected from maser emission associated 
with star formation regions, most of the masers are located within 
dusty, heavily obscured regions of the central starburst, particularly 
the masers located on the blue ward arc.


\subsection{Atomic and molecular gas}

\begin{figure*}
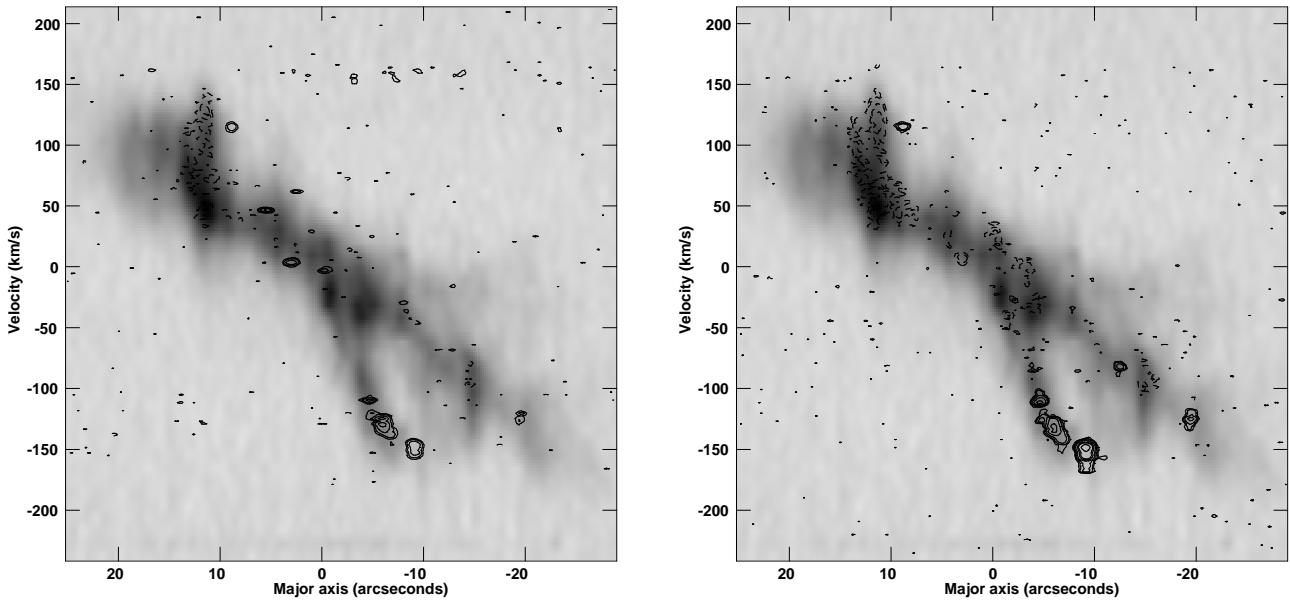

\begin{tabular}{cc}
\includegraphics[width=8.5cm]{1665+hi.ps}&
\includegraphics[width=8.5cm]{1667+hi.ps}\\
\end{tabular}
\caption{\label{oh+hi}A comparison between the OH and H{\sc i} 
datasets.  The 1665\,MHz $p-v$ plot is on the left, 1667\,MHz on the 
right.  The greyscale shows the H{\sc i} absorption while the 
contours show the OH absorption and emission.  The greyscale range is 
from 0 to $-$28 mJy/bm, while contours are drawn at 
($-$4,$-$2,$-$1,1,2,4,8,16,32,64) $\times$ 2.4\,mJy/bm (left) and 
1.5\,mJy/bm (right).  The centre of M82 is taken to be 09$^{\rm h}$55${\rm 
m}$52\rasec132 +69\degr40'46\farcs14 (J2000) and the velocity has been 
corrected for a systemic velocity of 225\,\kms.}
\end{figure*}

To carry out a comparison with the absorption seen in the atomic gas 
traced by H{\sc i} absorption, 1420\,MHz data from the VLA archive has 
been used from an A-array observation on November 27th 1996 (programme 
AW444; \citealt{wills00}).  The dataset was reduced using standard 
spectral line procedures in AIPS and a p-v plot was produced using the 
same method as described in Section \ref{section_velocity}.  While the 
spatial resolution of the two datasets is similar, the 1996 observation 
was carried out with a bandwidth of 3.125\,MHz over 63 channels 
resulting in a velocity resolution of 10.3\,\kms.  To compare this with 
the higher velocity resolution OH data from 2006, the two IFs at each 
frequency in the OH observations were combined using the AIPS task {\sc 
ujoin} resulting in two datasets of 190 channels each, one for each of 
the OH lines.  The H{\sc i} data has only 63 channels so to compare 
the datasets the number of channels was artificially increased using the 
AIPS task {\sc xsmth} which can interpolate over channels.  The 
resulting p-v plots are shown in Figure \ref{oh+hi}.

The deep absorption in the OH data is coincident with the deepest 
absorption features seen in the H{\sc i} data at the eastern end of the 
disk.  The masers appear to avoid the deepest H{\sc i} absorption 
features, but this could be due to OH absorption being present at the 
same locations since strong absorption can hide weak maser peaks (as 
illustrated by the 2002 VLA data).

The maser features on the blue arc are coincident with the arc seen in 
the H{\sc i} data on the blue ward side of the `hole' (\citealt{wills00})
also seen in the CO (\citealt{shen95}) and [Ne {\sc ii}] 
(\citealt{achtermann95}).  This feature is discussed further in Section 
\ref{bluearc}.
More recently, \cite{seaquist06} mapped M82 in $^{12}$CO J=6-5 and found 
evidence for the same blue feature, and a corresponding weak feature to 
the east of the galaxy, as well as higher than average excitation in the 
region surrounding 41.95+57.5 where the bright masers on this feature 
are located.

\subsection{Ionised gas}\label{ionised}

Comparing the positions of the masers with the [Ne{\sc ii}] emission 
from \cite{achtermann95}, several masers appear around the edge of the 
12.8$\mu$m ring.  The three bright features around the location of the 
radio continuum source 41.95+57.5 also appear to coincide with the 
[Ne{\sc ii}] peak labelled W1.  This is not surprising given that their 
velocity distribution also coincides with that of the [Ne{\sc ii}], 
rather than with the main disk rotation; the velocity distribution of 
the [Ne{\sc ii}] emission is much steeper than that of the main disk 
traced by the H{\sc i} or OH absorption and the western side of this 
steeper distribution is well traced by the brighter masers in this 
dataset.

\begin{figure*}
\includegraphics[width=8cm,angle=270]{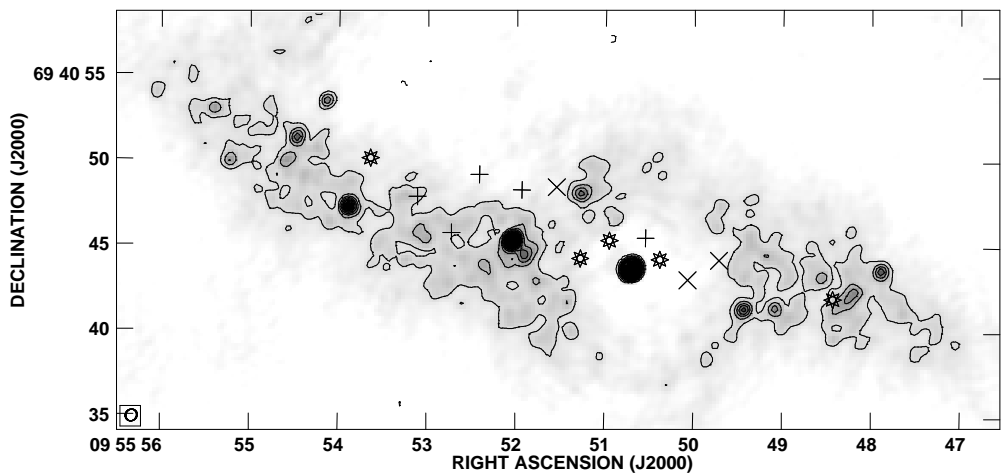}
\caption{\label{oh+408}Positions of the masers plotted over the 408\,MHz 
emission from MERLIN observations in 1994 from \protect\cite{wills97}.  
Contours are plotted at ($-$2, 2, 4, 6, 8, 10, 12, 14, 16, 18, 20) 
$\times$ 1\,mJy/bm.
As for Figure \ref{fighst}, masers visible at just 1665\,MHz, just 1667\,MHz, 
or both, are shown by plus, cross and star symbols repectively.   The 
masers avoid the areas of emission with the brightest masers sited in 
the extended absorption region surrounding the source 41.95+57.5, see 
Section \protect\ref{ionised}.}
\end{figure*}

Figure \ref{oh+408} shows the positions of the masers relative to the 
ionised gas as traced by 408\,MHz radio emission from a MERLIN 
observation in 1994 (\citealt{wills97}).  From this figure it can be 
seen that the masers tend to avoid the peaks of low frequency emission 
and, in particular, several of the masers are located within the `hole' 
feature noted by Wills et al.  This feature is $\sim$100\,pc in 
diameter, roughly centered on the compact object 41.95+57.5, and has 
been suggested to be due to free-free absorption by gas in a giant H{\sc 
ii} region ionised by a large stellar cluster (note that this is also 
the location of the high excitation seen in $^{12}$CO J=6-5 transition 
by \citealt{seaquist06} mentioned above).  The masers located in this 
region however are not typical of Galactic masers found in star forming 
regions as they all (except 50.54+45.5) have line ratios $>$1.  
Interestingly, it is also this group which are located on the blue 
feature in the p-v diagram except (again) 50.54+45.5 which has a 
velocity which places on the main distribution.

\subsection{The blue arc}\label{bluearc}

As has already been noted, an obvious feature of the p-v plots is the 
arc to the west of the disk which lies blue ward of the main 
distribution.  It has been shown previously that the OH absorption 
traces the arc well, and that several of the brightest maser features in 
M82 are located on the arc (\citealt{argo07}).
\cite{wills00} compare the distribution of H{\sc i} traced by absorption 
with that of CO emission and ionised [Ne{\sc ii}] finding a similar 
feature in both H{\sc i} and CO, and noted that the steeper gradient of 
the [Ne{\sc ii}] gas matches at least the eastern edge of this feature 
in the H{\sc i} and CO distributions.  This information was used to 
model a bar feature in the centre of M82.
\cite{matsushita05} showed a p-v plot of $^{12}$CO(1-0) together with 
H41$\alpha$ and noted that the known OH and H$_2$O masers were mainly 
clustered in velocity around the ``superbubble".
This arc is also present in recent optical data of emission lines over 
the central kpc of the disk (\citealt{westmoquette09}).

The gradient of the main distribution measured from these data is found 
to be $\sim$6\,km\,s$^{-1}$\,arcsec$^{-1}$, in agreement with 
$\sim$6.5\,\kms from H{\sc i} and CO measurements of \cite{wills00}, 
whereas the gradient of the emission on the eastern edge of the blue ward 
arc is $\sim$20\,km\,s$^{-1}$\,arcsec$^{-1}$, similar to that of the 
[Ne{\sc ii}] gas distribution.
The three brightest masers at 1667\,MHz are all located on this feature 
while the fainter features are on the main distribution expected for a 
rotating disk.  With the exception of 50.54+45.5, all of these masers 
are weaker at 1665\,MHz.

50.54+45.5 is the odd one out in this group since, despite appearing to 
be part of the group spatially, dynamically it is quite separate.  
Velocity measurements clearly place it not in the blue arc on the p-v 
diagram but on the main disk distribution tracing the bulk rotation of 
the galaxy.  It is also the only maser in this group which is visible 
only at 1665\,MHz, giving it a line ratio of $<$1 and, of the masers in 
this group, 50.54+45.5 lies closest to the edge of the dust lane in the 
optical map.

These observations confirm the results from \cite{argo07} that the 
brightest masers exist on this blue ward feature.  As well as being 
clustered in frequency, they are clustered spatially, in an apparent arc 
to the north of the strong continuum source 41.95+57.5.  Comparing the 
positions of the masers with optical emission in Figure \ref{fighst}, 
the masers in this group are located in a thick dust lane to the South 
of a bright knot.  They are also all spatially located around the 
[Ne{\sc ii}] features W1 and W2 from \cite{achtermann95} which shows a 
similar velocity distribution to that of the masers.

\section{Conclusions}

Sensitive, high spectral resolution observations of the M82 starburst 
have been carried out at 1.6\,GHz using the VLA in A-configuration.  The 
observations confirm previous detections and discovered three previously 
undetected masers.  All the masers are significantly brighter than 
typical Galactic masers, an effect which is probably due to the 
superposition of several masing clouds along a particular line of 
sight, a conclusion supported by the velocity structure we are 
starting to see in these higher resolution observations.

The spatial distribution of the masers follows that of the continuum 
emission with most of the masers apparently coincident with continuum 
objects and/or other maser species.  Their distribution in p-v space 
largely follows that of the absorption seen in H{\sc i}.  Most of the 
masers are located on the main velocity distribution which traces the 
bulk motion in the disk, however three of the brightest masers are 
located on a significant blue ward feature around a `hole' observed 
previously in H{\sc i} and CO, and traced particularly well by [Ne{\sc 
ii}] emission.
If this group of masers is tracing out an inner ring, as suggested by 
comparisons with the [Ne{\sc ii}] emission, then it might be expected 
that a corresponding group would be seen on the eastern side of the 
dynamical centre with redshifted velocities compared to the bulk 
distribution.  Such a feature is not traced either by maser emission or 
OH absorption, although a hint of such a structure was noted in H{\sc i} 
by \cite{wills00}.

Despite the improvement over previous observations, many of the masers 
are still largely unresolved in frequency.  However, those on the 
blue ward feature do show some interesting structures with multiple 
velocity components, supporting the idea that several maser features are 
being detected along one line of sight.  This structure is apparent at 
both 1665 and 1667\,MHz with all three masers more extended on the p-v 
diagram at 1667\,MHz.  50.95+45.4 shows particularly interesting 
structure with evidence of multiple components both spatially and in 
velocity.


\subsection*{Acknowledgements}

The VLA is operated by the National Radio Astronomy Observatory, a facility of the National Science Foundation operated under cooperative agreement by Associated Universities, Inc.
This research has made use of software provided by the UK's AstroGrid Virtual Observatory Project, which is funded by the Science and Technology Facilities Council and through the EU's Framework 6 programme.
RJB acknowledges support from the European Commission's I3 programme "RADIONET" under contract 505818.

\bibliographystyle{mn}
\bibliography{refs}

\begin{thebibliography}{23}
\expandafter\ifx\csname natexlab\endcsname\relax\def\natexlab#1{#1}\fi

\bibitem[{{Achtermann} \& {Lacy}(1995)}]{achtermann95}
{Achtermann} J.~M., {Lacy} J.~H., 1995, {Astrophys. J.}, 439, 163

\bibitem[{Argo {et~al.}(2007)Argo, Pedlar, Beswick, \& Muxlow}]{argo07}
Argo M.~K., Pedlar A.~P., Beswick R.~J., Muxlow T.~W.~B.~M., 2007, MNRAS, 380,
  596

\bibitem[{Baudry \& Brouillet(1996)}]{baudry96}
Baudry A., Brouillet N., 1996, {Astron. Astrophys.}, 316, 188

\bibitem[{{Beir{\~a}o} {et~al.}(2008){Beir{\~a}o}, {Brandl}, {Appleton},
  {Groves}, {Armus}, {F{\"o}rster Schreiber}, {Smith}, {Charmandaris}, \&
  {Houck}}]{beir08}
{Beir{\~a}o} P., {Brandl} B.~R., {Appleton} P.~N., {Groves} B., {Armus} L.,
  {F{\"o}rster Schreiber} N.~M., {Smith} J.~D., {Charmandaris} V., {Houck}
  J.~R., 2008, {Astrophys. J.}, 676, 304

\bibitem[{{Fenech} {et~al.}(2008){Fenech}, {Muxlow}, {Beswick}, {Pedlar}, \&
  {Argo}}]{fenech08}
{Fenech} D.~M., {Muxlow} T.~W.~B., {Beswick} R.~J., {Pedlar} A., {Argo} M.~K.,
  2008, {Mon. Not. R. Astr. Soc.}, 391, 1384

\bibitem[{{Freedman} {et~al.}(1994){Freedman}, {Hughes}, {Madore}, {Mould},
  {Lee}, {Stetson}, {Kennicutt}, {Turner}, {Ferrarese}, {Ford}, {Graham},
  {Hill}, {Hoessel}, {Huchra}, \& {Illingworth}}]{freedman94}
{Freedman} W.~L., {Hughes} S.~M., {Madore} B.~F., {Mould} J.~R., {Lee} M.~G.,
  {Stetson} P., {Kennicutt} R.~C., {Turner} A., {Ferrarese} L., {Ford} H.,
  {Graham} J.~A., {Hill} R., {Hoessel} J.~G., {Huchra} J., {Illingworth} G.~D.,
  1994, {Astrophys. J.}, 427, 628

\bibitem[{{Griffiths} {et~al.}(2000){Griffiths}, {Ptak}, {Feigelson},
  {Garmire}, {Townsley}, {Brandt}, {Sambruna}, \& {Bregman}}]{griffiths00}
{Griffiths} R.~E., {Ptak} A., {Feigelson} E.~D., {Garmire} G., {Townsley} L.,
  {Brandt} W.~N., {Sambruna} R., {Bregman} J.~N., 2000, Science, 290, 1325

\bibitem[{{Impellizzeri} {et~al.}(2008){Impellizzeri}, {McKean}, {Castangia},
  {Roy}, {Henkel}, {Brunthaler}, \& {Wucknitz}}]{impellizzeri08}
{Impellizzeri} C.~M.~V., {McKean} J.~P., {Castangia} P., {Roy} A.~L., {Henkel}
  C., {Brunthaler} A., {Wucknitz} O., 2008, {Nature}, 456, 927

\bibitem[{{Lo}(2005)}]{lo05}
{Lo} K.~Y., 2005, {Ann. Rev. Astron. Astrophys.}, 43, 625

\bibitem[{{Lonsdale}(2002)}]{lonsdale02}
{Lonsdale} C.~J., 2002, in IAU Symposium, Vol. 206, Cosmic Masers: From
  Proto-Stars to Black Holes, {Migenes} V., {Reid} M.~J., eds., pp. 413--+

\bibitem[{{Matsushita} {et~al.}(2005){Matsushita}, {Kawabe}, {Kohno},
  {Matsumoto}, {Tsuru}, \& {Vila-Vilar{\'o}}}]{matsushita05}
{Matsushita} S., {Kawabe} R., {Kohno} K., {Matsumoto} H., {Tsuru} T.~G.,
  {Vila-Vilar{\'o}} B., 2005, {Astrophys. J.}, 618, 712

\bibitem[{{McDonald} {et~al.}(2002){McDonald}, {Muxlow}, {Wills}, {Pedlar}, \&
  {Beswick}}]{mcdonald02}
{McDonald} A.~R., {Muxlow} T.~W.~B., {Wills} K.~A., {Pedlar} A., {Beswick}
  R.~J., 2002, {Mon. Not. R. Astr. Soc.}, 334, 912

\bibitem[{Muxlow {et~al.}(1994)Muxlow, Pedlar, Wilkinson, Axon, Sanders, \&
  de~Bruyn}]{muxlow94}
Muxlow T. W.~B., Pedlar A., Wilkinson P.~N., Axon D.~J., Sanders E.~M.,
  de~Bruyn A.~G., 1994, {Mon. Not. R. Astr. Soc.}, 266, 455

\bibitem[{{Rodriguez-Rico} {et~al.}(2004){Rodriguez-Rico}, {Viallefond},
  {Zhao}, {Goss}, \& {Anantharamaiah}}]{rodriguez04}
{Rodriguez-Rico} C.~A., {Viallefond} F., {Zhao} J.-H., {Goss} W.~M.,
  {Anantharamaiah} K.~R., 2004, {Astrophys. J.}, 616, 783

\bibitem[{Seaquist {et~al.}(1997)Seaquist, Frayer, \& Frail}]{seaquist97}
Seaquist E.~R., Frayer D.~T., Frail D.~A., 1997, {Astrophys. J.}, 487, L131

\bibitem[{{Seaquist} {et~al.}(2006){Seaquist}, {Lee}, \&
  {Moriarty-Schieven}}]{seaquist06}
{Seaquist} E.~R., {Lee} S.~W., {Moriarty-Schieven} G.~H., 2006, {Astrophys.
  J.}, 638, 148

\bibitem[{{Shen} \& {Lo}(1995)}]{shen95}
{Shen} J., {Lo} K.~Y., 1995, {Astrophys. J. {Letters}}, 445, L99

\bibitem[{Shopbell \& Bland-Hawthorn(1998)}]{shopbell98}
Shopbell P.~L., Bland-Hawthorn J., 1998, {Astrophys. J.}, 493, 129

\bibitem[{Weliachew {et~al.}(1984)Weliachew, Fomalont, \&
  Greisen}]{weliachew84}
Weliachew L., Fomalont E.~B., Greisen E.~W., 1984, {Astron. Astrophys.}, 137,
  355

\bibitem[{{Westmoquette} {et~al.}(2009){Westmoquette}, {Smith}, {Gallagher},
  {Trancho}, {Bastian}, \& {Konstantopoulos}}]{westmoquette09}
{Westmoquette} M.~S., {Smith} L.~J., {Gallagher} J.~S., {Trancho} G., {Bastian}
  N., {Konstantopoulos} I.~S., 2009, {Astrophys. J.}, 696, 192

\bibitem[{{Wills} {et~al.}(2000){Wills}, {Das}, {Pedlar}, {Muxlow}, \&
  {Robinson}}]{wills00}
{Wills} K.~A., {Das} M., {Pedlar} A., {Muxlow} T.~W.~B., {Robinson} T.~G.,
  2000, {Mon. Not. R. Astr. Soc.}, 316, 33

\bibitem[{{Wills} {et~al.}(1999){Wills}, {Pedlar}, {Muxlow}, \&
  {Stevens}}]{wills99}
{Wills} K.~A., {Pedlar} A., {Muxlow} T.~W.~B., {Stevens} I.~R., 1999, {Mon.
  Not. R. Astr. Soc.}, 305, 680

\bibitem[{{Wills} {et~al.}(1997){Wills}, {Pedlar}, {Muxlow}, \&
  {Wilkinson}}]{wills97}
{Wills} K.~A., {Pedlar} A., {Muxlow} T.~W.~B., {Wilkinson} P.~N., 1997, {Mon.
  Not. R. Astr. Soc.}, 291, 517

\end{thebibliography}

\end{document}